\documentclass[%
 reprint,
superscriptaddress,
 amsmath,amssymb,
 aps,
 pra,
longbibliography
]{revtex4-1}

\usepackage{lipsum}
\usepackage{xcolor}
\usepackage{graphicx}% Include figure files
\usepackage{dcolumn}% Align table columns on decimal point
\usepackage{bm}% bold math
\usepackage{comment}
\usepackage{soul}
\usepackage{hyperref}% add hypertext capabilities
\usepackage{qcircuit}
\usepackage{tikz}
\usepackage{braket}
\usepackage{soul} % for strikeouts

\newcommand{\bs}[1]{\boldsymbol{#1}}

\begin{document}

\preprint{APS/123-QED}

\title{
Gauge invariant quantum circuits for $U(1)$ and Yang-Mills lattice gauge theories
}

\author{Giulia Mazzola}
\affiliation{IBM Quantum, IBM Research – Zurich, 8803 Rüschlikon, Switzerland}
\affiliation{Institute for Theoretical Physics, ETH Zürich, 8093 Zürich, Switzerland}

\author{Simon V. Mathis}
\affiliation{IBM Quantum, IBM Research – Zurich, 8803 Rüschlikon, Switzerland}

\author{Guglielmo Mazzola}
\affiliation{IBM Quantum, IBM Research – Zurich, 8803 Rüschlikon, Switzerland}

\author{Ivano Tavernelli}
\affiliation{IBM Quantum, IBM Research – Zurich, 8803 Rüschlikon, Switzerland}

\date{\today}% It is always \today, today,

\begin{abstract}
Quantum computation represents an emerging framework to solve lattice gauge theories (LGT) with arbitrary gauge groups, a general and long-standing problem in computational physics. 
While quantum computers may encode LGT using only polynomially increasing resources, a major open-issue concerns the violation of gauge-invariance during the  dynamics and  the search for ground-states.
Here, we propose a new class of parametrized quantum circuits that can represent states belonging only to the physical sector of the total Hilbert space.
This class of circuits is compact yet flexible enough to be used as a variational ansatz to study  ground state properties, as well as representing states originating from a real-time dynamics.
Concerning the first application, the structure of the wavefunction ansatz guarantees the preservation of physical constraints such as the Gauss law along the entire optimization process, enabling reliable variational calculations.
As for the second application, this class of quantum circuits can be used in combination with time-dependent variational quantum algorithms, thus drastically reducing the resource requirements to access dynamical properties.

\end{abstract}

\pacs{Valid PACS appear here}
\maketitle

\section{Introduction} 
Gauge theories lie at the heart of the Standard Model of particle physics and represent the most successful description of elementary particles and their fundamental interactions~\cite{YangMills1954,Peskin1995}. For instance, these theories represent a generalization of quantum electrodynamics (QED) which elucidates the behavior of charged particles and photons, and quantum chromodynamics (QCD) which describes the strong interactions between quarks, the elementary constituents of protons
and neutrons.
%In the past few decades, the predictions of Yang-Mills gauge theories have shown an excellent agreement with experimental data~\cite{Altarelli2017}. 
While perturbative approaches exist in the study of gauge theories, for example describing weak or electromagnetic interactions~\cite{Weinberg1995vol2}, other theories can be approached with these methods only in certain limiting regimes. For instance, QCD can be studied perturbatively in the limit of high energy while in the low-energy regime however, the strong interactions grow so large that a perturbative analysis is not possible anymore.

A crucial step towards a non-perturbative analysis of gauge theories such as QCD was taken by Wilson who formulated gauge theories on a finitely discretized space-time lattice while preserving the exact local symmetry of a gauge theory, thereby introducing the concept of lattice gauge theories (LGT)~\cite{Wilson1974}.
Despite the success of path-integral Monte Carlo sampling methods and tensor network approaches~\cite{creutz1983monte,Byrnes2002, Buyens2014,Silvi2014, Silvi2017, Silvi2019,Pichler2016,Dalmonte2016,PhysRevLett.117.182001,Montangero_book,ZOHAR2015-Ann,Zohar2018, Robaina2021} for estimating equilibrium properties of LGT~\cite{PP-review-2018_short}, out-of-equilibrium and real-time dynamics have remained out of reach, as they are limited by a severe numerical sign problem that exponentially increases the time complexity with increasing system size~\cite{Troyer2005}, as well as by the exponential growth of the entanglement with the simulation time in arbitrary dimensions~\cite{Banuls2017}. 

Quantum computing offers a promising alternative, as it does not suffer from the above limitations. By expressing LGTs in the equivalent Hamiltonian formalism~\cite{KogutSusskind1975}, quantum simulators may be built to simulate the LGT dynamics~\cite{Banuls2019SimulatingLG,Zohar2011, Zohar2013-PRL, PhysRevResearch.2.013288,PhysRevD.95.094507,PhysRevLett.115.240502,yang2020observation,Byrnes2006, Andrist_2011,Banerjee2012,Banerjee2013,Tagliacozzo2013160,Tagliacozzo2013,Yang2016,Martinez2016,Kokail2019,Mathis2020,PhysRevD.100.034518,Mil2020} with resources that only grow polynomially with the system size.
However, one main problem that arises in the standard Hamiltonian formulation of a LGT~\cite{KogutSusskind1975} is the emergence of (exponentially many) \emph{unphysical} states~\cite{Mathis2020,PhysRevLett.125.030503,PhysRevResearch.2.033361,yang2020observation}.  
The origin of such an unphysical subspace is related to the additional freedom associated to a particular choice of the gauge fixing implicit to the canonical quantization of the Hamiltonian. In the case of the temporal (partial) gauge, the physical subspace is defined by additional so-called \emph{Gauss law} constraints.
For instance, in the case of QED the Gauss law takes the familiar form $\nabla\cdot\bs{E} = -e\, \psi^\dagger \psi$ which relates the electric charge distribution $-e\, \psi^\dagger \psi$ to the electric field $\bs{E}$. As a consequence, such constraints resulting from the gauge symmetry of LGT must be incorporated either into the encoding of the relevant degrees of freedom or into the operators which govern the evolution of the system.

%Proposals exist to either (i) integrate these constraints directly into the Hamiltonian~\cite{Bringoltz2009, Martinez2016, Sala2018, Zohar2019, Kaplan2020, Bender2020}, thereby effectively erasing the gauge or matter degrees of freedom, (ii) incorporate them into the encoding of the states~\cite{Stryker2019, Felser2020}, or (iii) implement them as a penalty function in the Hamiltonian~\cite{Zohar2011,lacroix2011introduction,Dalmonte2016,Mathis2020}. However, these approaches come at the cost of losing the locality of the Hamiltonian or of restricting their applicability to specific models and qubit encoding schemes, making their generalization to other gauge groups or to higher space dimensions impossible.}

In this manuscript, we investigate the task of state preparation while retaining the gauge symmetry constraints imposed by LGT as mentioned above.
State preparation is the first step in all quantum algorithms targeting ground state calculations, ranging from the variational quantum eigensolver (VQE)~\cite{Peruzzo2014} to quantum phase estimation (QPE)~\cite{nielsen_chuang_2010,abrams1999quantum}, thus representing a key stage in virtually every kind of quantum simulation.
State preparation is also crucial for the initialization of non-equilibrium processes (assuming  that  the  initial  state of interest is not trivially represented in the computational basis) originating from a quench~\cite{schweizer2019floquet}.

To this end, we introduce a class of parametrized quantum circuits that realize states that are gauge invariant by construction and applicable to generic Yang-Mills LGT theories in any spacetime dimension. 
Notably, the proposed approach is independent of the choice of a specific qubit encoding. We observe a clear advantage over an unstructured variational form ansatz, both in the estimation of ground state wavefunctions as well as in the implementation of real time dynamics. 

%%%%%%%%%%%%%%%%%%%%%%%%%%%%%%%%%%%%%%%%%%%%%%%%%%%%%

%%%%%%%%%%%%%%%%%%%%%%%%%%%%%%%%%%%%%%%%%%%%%%%%%%%%%
%%%%%%%%%%%%%%%%%%%%%%%%%%%%%%%%%%%%%%%%%%%%%%%%%%%%%
%%%% FIGURE 3 YM SETUP
\begin{figure*}[hbt]
\includegraphics[width=0.85\textwidth]{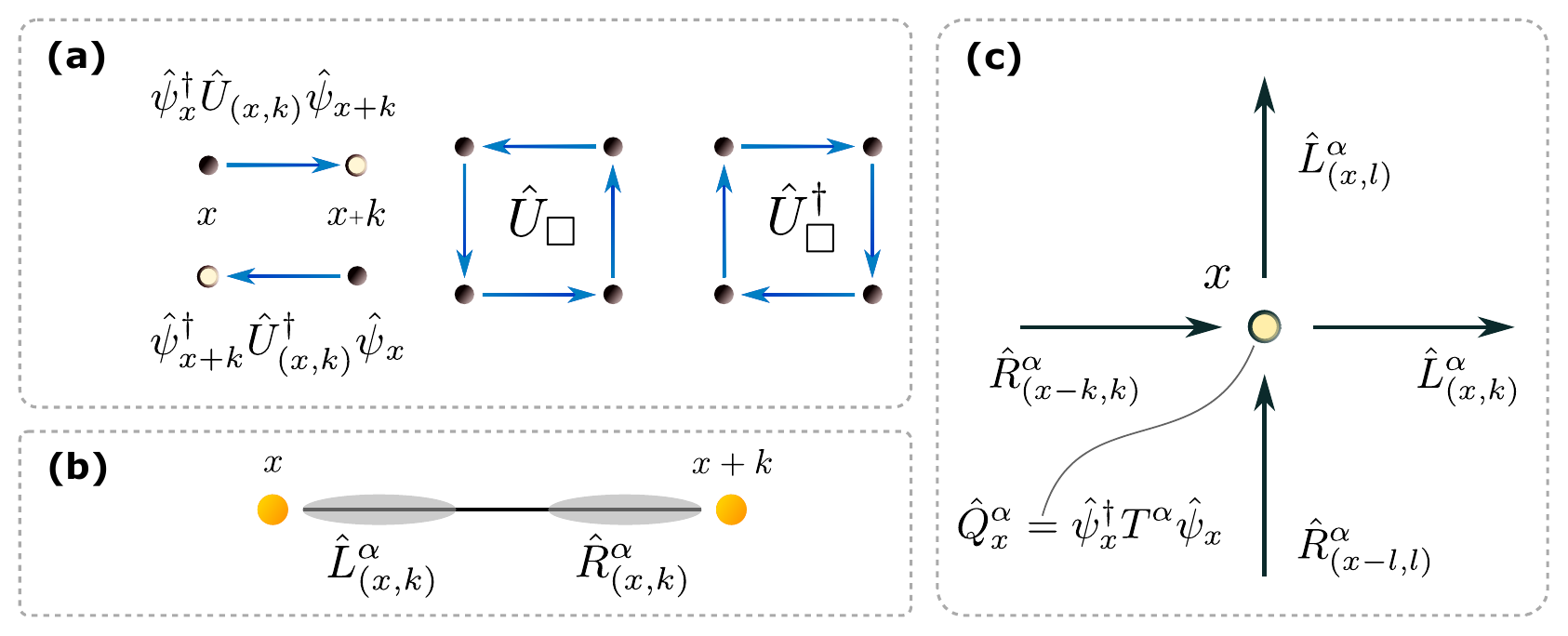}
\caption{Graphical illustration of the components describing a Yang-Mills model on a lattice. \textbf{(a)} Visualization of the hopping terms $\hat{\psi}^\dagger \hat{U} \hat{\psi}$ and the plaquette terms $\hat{U}_\Box$ plus their hermitian conjugates on a lattice. \textbf{(b)} Illustration of a non-Abelian link. The colored left operators $\hat{L}^\alpha_{(x,k)}$ and right operators $\hat{R}^\alpha_{(x,k)}$ are independent variables that are assigned to the left end and the right end of a link $(x,k)$, respectively. \textbf{(c)} Visualization of the non-Abelian Gauss law constraint. The difference between the outgoing flux and the incoming flux on a site $x$ equals the amount of charge $\hat{Q}$ that sits on that site.}
\label{fig:ym-model}
\end{figure*}
%%%%%%%%%%%%%%%%%%%%%%%%%%%%%%%%%%%%%%%%%%%%%%%%%%%%%
%%%%%%%%%%%%%%%%%%%%%%%%%%%%%%%%%%%%%%%%%%%%%%%%%%%%%

\section{Yang-Mills Model Hamiltonian} We start by introducing a lattice model that reproduces the dynamics of a Yang-Mills gauge theory with dynamical fermionic matter in the continuum limit~\cite{KogutSusskind1975, Zohar2013-PRA, Wiese2013, Zohar2015, Mathis2020}. In Yang-Mills gauge theories, multiple matter particle species $\psi_a$ (representing different types of quarks, for example) can exist such that the theory is invariant under certain local transformations $V(\bs{x})$ that might mix the different particle species $\psi_a(\bs{x}) \rightarrow  V_{ab}(\bs{x})\psi_b(\bs{x})$ at every space-time point $\bs{x}$. These local transformations are characterized by an Abelian or non-Abelian compact Lie group $G$, the so-called the \emph{gauge group} of the theory~\cite{Peskin1995}. More specifically, the transformations $V(\bs{x})$ correspond to a finite unitary representation of the gauge group $G$.
In this setting, QED can be regarded as a Yang-Mills theory with the Abelian gauge group $U(1)$ with one single particle species representing the electron, while QCD corresponds to a Yang-Mills theory with the non-Abelian gauge group $SU(3)$ describing three particle species which represent the \emph{colors} of quarks. 

In this work, the Yang-Mills theory is formulated on a $d$-dimensional discrete spatial lattice with lattice spacing $a$. 
%The time component is kept as a continuous parameter.
We denote lattice sites in $d$ dimensions by their coordinate vector $x \in \mathbb{R}^d$ and edges (also called links) between sites by the tuple $(x, k)$ where $x$ is a lattice site and $k \in \lbrace 1, ... d \rbrace $ is a direction. The (single flavor) fermionic particles reside on the lattice sites $x$, while the gauge fields live on the links. The Yang-Mills Hamiltonian $\hat{H}_\text{YM} = \hat{H}_\text{hopp} \, + \, \hat{H}_\text{mass}  + \hat{H}_\text{plaq} + \hat{H}_\text{elec} + \hat{H}_\text{wilson}$ in the temporal gauge reads
\begin{align}\label{eq:final-YM-hamiltonian}
    \hat{H}_\text{YM} &= \sum_{\text{sites}} \sum_k \frac{1}{2a} \left( \hat{\bar{\psi}}_x \left[ i\gamma^k + r\right]\hat{U}_{(x,k)}\hat{\psi}_{x+ k} + \text{h.c.} \right) \nonumber \\
    +& \sum_\text{sites} \left( m + \frac{rd}{a} \right)\hat{\bar{\psi}}_x \hat{\psi}_x \,-\, \frac{a^{d-4}}{2g^2}\sum_\text{plaq.} \text{Tr}\left( \hat{U}_\square + \hat{U}^\dagger_\square \right) \nonumber \\
    +& \,\frac{g^2a^{2-d}}{4} \sum_\text{links}\sum_\alpha \big((\hat{L}^{\alpha}_{(x,k)})^2 + (\hat{R}^{\alpha}_{(x,k)})^2 + \theta^\alpha_k\, \big)\,,
\end{align}
where $\gamma^k$ are given by the Dirac gamma matrices and where $m$, $g$ denote the fermionic mass parameter and the gauge field coupling parameter, respectively. This Hamiltonian mainly consists of four contributing terms describing the fermionic matter ($\hat{H}_\text{mass}$) and gauge field energies ($\hat{H}_\text{elec}$ and $\hat{H}_\text{plaq}$) and their interactions ($\hat{H}_\text{hopp}$). An additional term $\hat{H}_\text{wilson}$ is introduced via the parameter $r$ which acts as a regulator to avoid the fermion doubling problem~\cite{Kogut1983, Zache2018}. The quantities $\hat{\psi}_x$ and $\hat{\bar{\psi}}_x = \hat{\psi}_x^\dagger \gamma^0$ represent the fermionic field operators that carry implicit spinor- and color component indices $ i,j \,\in \left\{1,\dots, 2^{\lceil d/2 \rceil}\right\}$ and $b,c \,\in \left\{1,\dots, \dim(R) \right\}$ respectively, where $\dim(R)$ denotes the dimension of the representation $R$ of the corresponding gauge group $G$. For example, in the case of $G=SU(N)$ in the fundamental representation, we have $\dim(R) = N$. Note that whenever the spinor- and color indices are suppressed, we implicitly sum over all these indices. The fermionic operators satisfy the standard anti-commutation relations
\begin{equation}
    \big\{ \hat\psi^{b,i}_x, \hat\psi^{\dagger c,j}_y \big\} = \delta_{xy}\delta_{bc}\delta_{ij}, \ \big\{\hat\psi^{b,i}_x, \hat\psi^{c,j}_y \big\} = 0 \,.
\end{equation}
The gauge field operators $\hat{U}_{(x,k)}$ on a link $(x,k)$ have the same structure as the generators $T^\alpha$ with $\alpha \in \{ 1,\dots, \dim (G)\}\,$ of the gauge group: They are $\dim(R) \times \dim(R)$ dimensional matrices whose entries $(\hat{U}_{(x,k)})_{mn}$ are operators that act on the link Hilbert space. The plaquette operators are defined by $ \hat{U}_\square \, =\,  \hat U_{(x,k)} \hat U_{(x+k,l)}\hat U^\dagger_{(x+l,k)}\hat U^\dagger_{(x,l)} \,$, and the trace of these operators appearing in $\hat{H}_\text{YM}$ is only taken in the Lie algebra space of the group generators $T^\alpha$ and does not extend to the quantum Hilbert space.
The conjugate link flux variables $\hat{L}^\alpha_{(x,k)}$ and $\hat{R}^\alpha_{(x,k)}$ are associated with the left end and the right end of a link $(x,k)$, respectively, and are defined through the commutation relations~\cite{Brower1999}
\begin{equation}
    \big[ \hat{L}^\alpha_{(x,k)}, \hat{U}_{(x,k)}\big] =\, T^\alpha\hat{U}_{(x,k)}\,, 
    \big[ \hat{R}^\alpha_{(x,k)},\hat{U}_{(x,k)}\big] =\, \hat{U}_{(x,k)}T^\alpha\,.
\end{equation}
Note that in the case of QED, these two variables reduce to the electric flux field operators $\hat{E}_{(x,k)}$ used in~\cite{Mathis2020}. In addition, a constant background gauge field is introduced via the constant parameters $\theta^\alpha_k$~\cite{Funcke2020}. 
Note that all operators assigned to different lattice sites or links commute with each other. 
The Gauss law operators are defined by
\begin{equation}\label{eq:gauss-law-operators}
    \hat{G}^\alpha_x \,=\, \sum_k \left(\hat{L}^\alpha_{(x,k)} - \hat{R}^\alpha_{(x-k,k)}\right) - \hat{\psi}^\dagger_x T^\alpha \hat{\psi}_x
\end{equation}
and they specify the gauge-invariant states $|\phi_\text{phys}\rangle$ that span the physical sector $\mathcal{H}^\text{phys}$ of the total Hilbert space via 
\begin{equation}
    |\phi_\text{phys}\rangle \,\in\, \mathcal{H}^\text{phys} \quad \Leftrightarrow \quad \hat G^\alpha_x|\phi_\text{phys}\rangle \,=\, 0\,
\end{equation}
for all lattice sites $x$ and group generator indices $\alpha$.
The Gauss law operators are the generators of local (time-independent) gauge transformations and commute with the Hamiltonian $\big[\hat{H}_\text{YM}, \hat{G}^\alpha_x\big] \,=\,0$, implying that the gauge-invariance is in principle conserved in a real-time evolution.
Finally, the Hilbert space $\mathcal{H}_\text{YM}$ on which the Hamilton operator $\hat{H}_\text{YM}$ acts is given by $ \mathcal{H}_\text{YM}\,=\, \bigotimes_{x,k}\, \mathcal{H}^\text{fermi}_x \otimes \mathcal{H}^\text{gauge}_{(x,k)}$, where the individual (fermionic) Hilbert spaces at each lattice site $x$ have dimension $\dim(\mathcal{H}^\text{fermi}_x)\,=\, 2^{2^{\lceil d/2 \rceil}\dim(R)} $, while the dimension of the (bosonic) link Hilbert spaces $\mathcal{H}^\text{gauge}_{(x,k)}$ generally depends on a chosen truncation of the gauge fields~\cite{Orland1990, Brower1999, Byrnes2006} since otherwise, it would be infinite dimensional due to the continuous nature of the gauge group. The physical states $|\phi_\text{phys}\rangle$ which preserve the Gauss law constraints span the physical Hilbert subspace $\mathcal{H}^\text{phys} \subset \mathcal{H}_\text{YM}$. The various components of the Yang-Mills model Hamiltonian are visualized in Fig~\ref{fig:ym-model}.

\section{Enforcing the Gauss law constraints}

In the standard Hamiltonian formulation of Yang-Mills LGT, we have seen that the model Hamiltonian~\eqref{eq:final-YM-hamiltonian} operates on a large Hilbert space $\mathcal{H}_\text{YM}$ containing all possible matter and gauge field states, especially including the unphysical states, which are not gauge-invariant. When studying the spectrum of the Yang-Mills Hamiltonian using for instance a variational approach, one has to remain within the physical subspace $\mathcal{H}^\text{phys}$ with gauge-invariant states $|\phi_\text{phys}\rangle$ in order to guarantee the correct physical energy spectrum. On the other hand, in the case of real-time evolution initiated from a gauge invariant state $|\phi_\text{phys}(0)\rangle$, the time-evolved state $|\phi_\text{phys}(t)\rangle$ at time $t$ will remain gauge invariant since the Gauss law constraints are constants of motion, $[\hat{G}^\alpha_x, \hat{H}_\text{YM}]=0$. For this reason, there would in principle be no need to additionally enforce the Gauss law. 
Nevertheless, errors that kick the state out of the physical Hilbert space $\mathcal{H}^\text{phys}$ can occur due to the Trotter approximation~\cite{Trotter1959, Suzuki1976} of the time-evolution operator or from quantum hardware noise.

Various implementations of the Gauss law constraints have been proposed in the literature, 
ranging from absorbing these constraints directly into the Hamiltonian~\cite{Bringoltz2009, Martinez2016, Sala2018, Zohar2019, Kaplan2020, Bender2020} (thereby effectively removing the explicit gauge or matter degrees of freedom), to their incorporation into the state ansatz (gauge and matter particle)~\cite{Stryker2019, Felser2020}.
%for example to integrate these constraints directly into the Hamiltonian~\cite{Bringoltz2009, Martinez2016, Sala2018, Zohar2019, Kaplan2020, Bender2020}, thereby effectively erasing the gauge or matter degrees of freedom, or to incorporate them into the encoding of the (gauge and matter particle) states~\cite{Stryker2019, Felser2020}. 
%These approaches usually are specific to a particular setups or model dimensions, or more suitable for analog quantum simulators and therefore not always convenient in digital simulations.
However, these approaches are often limited to specific models and particular dimensions and are not of general applicability for digital quantum simulations.

A more pragmatic approach consists in enforcing the Gauss law constraints by means of an energy penalty term
\begin{equation}
    \hat{H}_\text{gauge} = \lambda\sum_{x,\alpha} (\hat{G}^\alpha_x)^2
\end{equation} 
added to the system Hamiltonian and proportional to the regularization parameter $\lambda$~\cite{Zohar2011,lacroix2011introduction,Dalmonte2016,Mathis2020}. 
This term effectively lifts the energy of unphysical states while the physical gauge-invariant states, lying in the kernel of the Gauss law operators, remain unaffected. As a result, for large enough values of $\lambda$, the low-lying energy spectrum is solely associated with physical states. 
It is worth mentioning that modifications of this approach exist in which the penalty term is replaced by simpler terms that become linear in the Gauss law operators~\cite{Stannigel2014, Halimeh2020, Kasper2021}.

This approach is general enough to be applied regardless of the lattice dimensionality, the gauge group and the qubit encoding. 
However, the penalty term implements the Gauss law constraints only approximately, requiring a tuning of the parameter $\lambda$, and, in addition, they come at the expense of increasing the complexity of the Hamiltonian. 
In the context of variational quantum algorithms for groundstate optimization, we further argue in Appendix~\ref{app:performance-vqe} and~\ref{app:1d-example_sampling} that $\hat{H}_\text{gauge}$ leads to a more corrugated energy landscape (see for instance~\cite{Berthier_2009})
% (since the energy of the unphysical states is significantly lifted compared to the scales of the Hamiltonian parameters), 
and that a large regularization parameter $\lambda$ induces significant sampling errors in estimating the energy expectation value, thereby aggravating the optimization process in the variational algorithm. 

The above considerations thus motivate the preparation of states for which the Gauss law constraints need not be additionally enforced but are automatically fulfilled by construction.

%%%%%%%%%%%%%%%%%%%%%%%%%%%%%%%%%%%%%%%%%%%%%%%%%%%%%
\section{Gauge-invariant state preparation}
Heuristic state preparation ans\"{a}tze~\cite{Kokail2019}, which are common for example in quantum chemistry applications~\cite{McClean2016,Kandala2017},  fail to sample efficiently from $\mathcal{H}^\text{phys}$ (see Appendix~\ref{app:performance-vqe}). 
To circumvent this problem, we construct parametrized trial states tailored to the gauge symmetry of the theory. To do so, we look for unitary operators $\hat{\mathcal{O}}$ that map the physical Hilbert space $\mathcal{H}^\text{phys}$ onto itself, i.e. $\hat{\mathcal{O}}\ket{\phi_\text{phys}} \in \mathcal{H}^\text{phys}$ for all $\ket{\phi_\text{phys}} \in \mathcal{H}^\text{phys}$. We call unitary operators with this property \emph{gauge invariant}. A parametrized family $\hat{\mathcal{U}}(\bs{\theta})$, with real parameter vector $\bs{\theta}$, of gauge invariant operators then allows us to construct trial states from any initial state $\ket{\phi_\text{init}} \in \mathcal{H}^\text{phys}$ via
\begin{equation}
    \ket{\phi(\bs{\theta})} = \hat{\mathcal{U}}(\bs{\theta})\ket{\phi_\text{init}}\,.
\end{equation} 
The definition of $\mathcal{H}^\text{phys}$ implies that a family of unitary operators $ \hat{\mathcal{U}}(\bs{\theta})$ is gauge invariant if and only if it commutes with the Gauss law operators $\hat{G}^\alpha_x$ in Eq.~\eqref{eq:gauss-law-operators},
$
    \big[\hat{\mathcal{U}}(\bs{\theta}), \hat{G}^\alpha_x \big] \,=\, 0 \,
$
 for all $\bs{\theta},\alpha,x\,$.
With this choice, it follows that a parametrized trial state $\ket{\phi(\bs{\theta})}$ is gauge invariant if and only if the initial state $\ket{\phi_\text{init}}$ is chosen to be gauge invariant since $ \hat{G}^\alpha_x\ket{\phi(\bs{\theta})} \,=\, \hat{\mathcal{U}}(\bs{\theta})\,\hat{G}^\alpha_x\ket{\phi_\text{init}} \,=\, 0 \,$. As a result, such a gauge-invariant variational form will sample from the physical Hilbert space only. Furthermore, note that due to the linearity and the product rule $[AB, C] = A[B,C] + [A,C]B$ of the commutator for arbitrary operators $A$, $B$ and $C$, it follows that any linear combination or product of gauge invariant terms will remain gauge invariant. Thus, if a Hermitian operator commutes with the Gauss law operators $\big[\hat{F}, \hat{G}^\alpha_x  \big]=0$, then the exponential of $\hat{F}$ is a unitary operator that commutes with the Gauss law operators, $\big[\exp(i\hat{F}), \hat{G}^\alpha_x  \big]=0$, too.

Following these prescriptions, a parametrized family of gauge-invariant unitary operators $\hat{\mathcal{U}}(\bs{\theta})$ can be generated by combining pieces of the Yang-Mills Hamiltonian $\hat{H}_\text{YM}$ in Eq.~\eqref{eq:final-YM-hamiltonian} which commute with the Gauss law operators. These pieces can be promoted to the following parametrized terms:
\begin{align}
    &\text{Mass-like}  &\hat{\psi}^\dagger_x\,A(\bs{\theta})\,\hat{\psi}_x  \label{eq:variational-form-toolkit-1} \\
    &\text{Left-hopping-like}  & \hat{\psi}^\dagger_x\,A(\bs{\theta})\,\hat{U}_{(x,k)}\,\hat{\psi}_{x+k}  \label{eq:variational-form-toolkit-2}\\
    &\text{Right-hopping-like}  & \hat{\psi}^\dagger_{x+k}\,A(\bs{\theta})\,\hat{U}^\dagger_{(x,k)}\,\hat{\psi}_x  \label{eq:variational-form-toolkit-3} \\
    &\text{Left-flux-like}  & \Lambda(\bs{\theta})\,\sum_\alpha \big(\hat{L}^\alpha_{(x,k)} \big)^2 \label{eq:variational-form-toolkit-4}\\
    &\text{Right-flux-like}  & \Lambda(\bs{\theta})\,\sum_\alpha \big(\hat{R}^\alpha_{(x,k)} \big)^2 \label{eq:variational-form-toolkit-5}\\
    &\text{String-like}  & \hat{\psi}^\dagger_x\,A(\bs{\theta})\,\hat{U}_1\hat{U}_2\cdots\hat{U}_l\,\hat{\psi}_{x+l} \label{eq:variational-form-toolkit-6}\\ 
    &\text{Loop-like}  & \Lambda(\bs{\theta})\,\text{Tr}\bigg(\,\prod_{l\in \mathcal{C}}\,\hat{U}_{l}\bigg) \label{eq:variational-form-toolkit-7}
\end{align}
where 
$\Lambda(\bs{\theta})$ is a real scalar, and
the \textit{coupling matrix} $A(\bs{\theta})$ is an arbitrary $n_\text{spinor}\times n_\text{spinor}$ matrix that mixes the spinor components of the fermionic matter fields. We allow the mixing of the spinor components since these spinors are all affected in the same manner under a gauge transformation and a mixing does not change the gauge invariance properties of these terms. On the other hand, mixing the color components in an arbitrary manner would destroy the gauge symmetry of these terms. Note that we implicitly sum over all color and spinor indices $b,c$ and $i,j$. 
For example, Eq.~\eqref{eq:variational-form-toolkit-2} would more explicitly read $\sum_{i,j,b,c} (\hat{\psi}^\dagger_x)_{b,i}\,A(\bs{\theta})_{ij}\,\big(\hat{U}_{(x,k)}\big)_{bc}\,(\hat{\psi}_{x+k})_{c,j} $. 
In the \emph{string-like} term, the link variables $\hat{U}_n$ can be chosen such that their product forms an arbitrary path which connects the lattice sites $x$ and $x+l$ along lattice edges. The set $\mathcal{C}$ in the loop-like term denotes an ordered sequence of links that form a loop.
We then construct a gauge invariant unitary variational form $\hat{\mathcal{U}}(\bs{\theta})$ by defining 
\begin{equation}
    \hat{\mathcal{U}}(\bs{\theta}) \,=\, \prod_k \exp\left(i\,\big[\hat{F}_k(\bs\theta) + \hat{F}_k^\dagger(\bs\theta) \big]\right)\,,
\end{equation}
where $\hat{F}_k(\bs\theta)$ denotes any linear combination or product of the gauge invariant terms~\eqref{eq:variational-form-toolkit-1} --~\eqref{eq:variational-form-toolkit-7}.
In this respect, the present variational form is conceptually different from the so-called Hamiltonian variational ansatz introduced in~\cite{wecker2015progress}, inspired by a trotterization of an annealing process.

Conceptually, the set of operators~\eqref{eq:variational-form-toolkit-1} --~\eqref{eq:variational-form-toolkit-7} is to be considered as a toolbox for preparing generic gauge-invariant states in a particular state space $\mathcal{H}\subset \mathcal{H}^\text{phys} $. For instance, considering a one-dimensional QED system in a low-energy regime, one might expect large flux strings to be absent (as the energy scales with the length of the flux string) and thus, one would not include any string-like term but rather only (nearest-neighbor) hopping-like terms in the construction of the variational form, as we will see in the following section. Finally, note that since this set actually contains all terms appearing in the Hamiltonian of the system, it is reasonable to expect~\cite{wecker2015progress} that these pieces are sufficient to represent the whole physical Hilbert space of states for a particular choice of the variational parameters.

%%%%%%%%%%%%%%%%%%%%%%%%%%%%%%%%%%%%%%%%%%%%%%%%%%%%%
%%%%%%%%%%%%%%%%%%%%%%%%%%%%%%%%%%%%%%%%%%%%%%%%%%%%%
%%%% FIGURE 1 VQE
\begin{figure*}[hbt]
\includegraphics[width=0.95\textwidth]{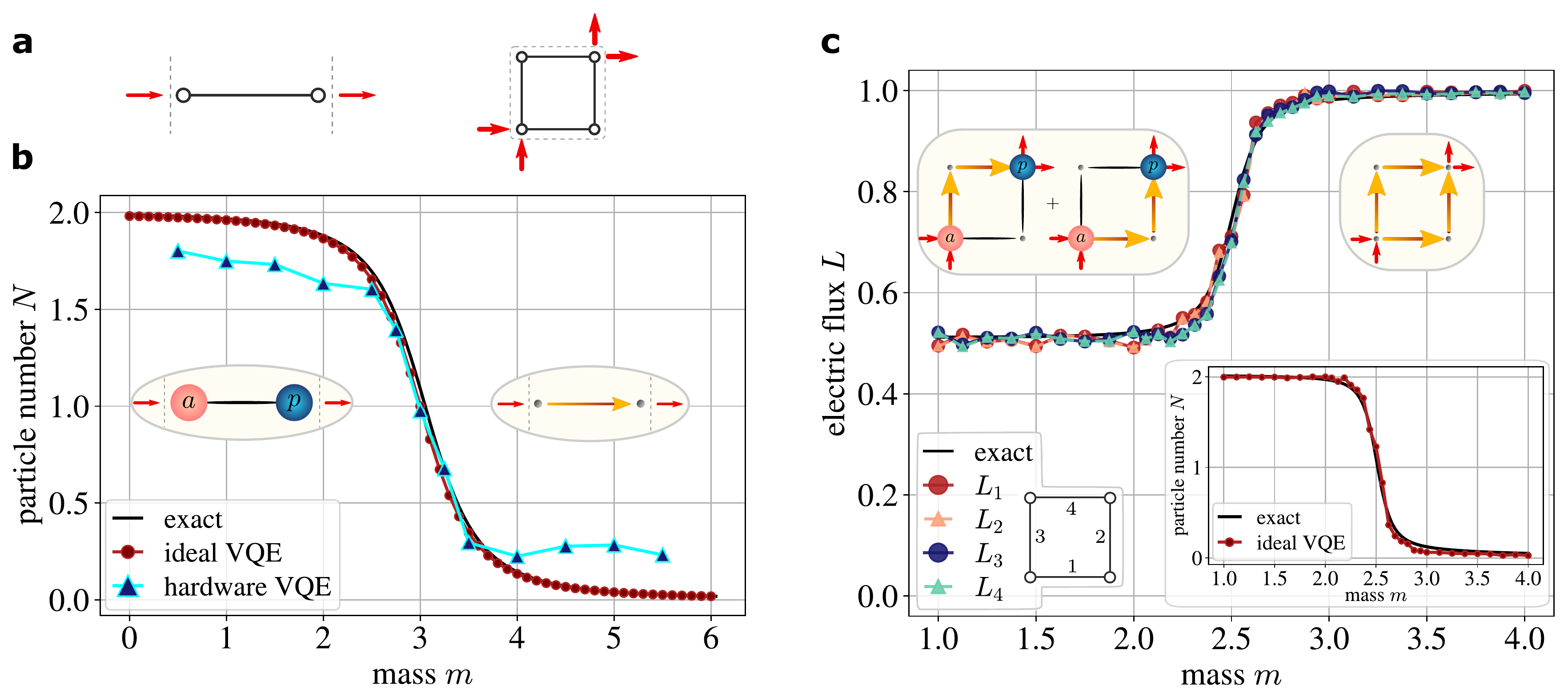}
\caption{Ground state search in a (1+1)- and (2+1)-dimensional lattice QED model using a VQE algorithm. \textbf{a)} Lattice configurations in $d=1$ (left) and $d=2$ (right) space dimensions with boundary conditions as indicated by red arrows. Such conditions might be induced by (anti-)particles sitting outside of the lattice. \textbf{b)} The particle number $N$ of the ground state is shown as a function of varying mass $m$ for $d=1$, resulting from an ideal VQE algorithm (red dots) and from a hardware simulation on a superconducting quantum device (blue triangles). The resources reduce to 5 qubits, 16 cnots and 3 variational parameters. We observe the string-breaking phenomenon in a form reminiscent of a phase transition from a particle-antiparticle pair (left inset) to a flux-tube configuration (right inset) at critical mass $m_c =3$.  \textbf{c)} The electric flux $L_i$ for each lattice link $i$ and the particle number $N$ (lower-right inset) of the ground state are plotted as a function of varying mass $m$ for $d=2$, resulting from an ideal VQE algorithm. Only one exact curve (black) for the electric flux is shown since the ground state flux configurations on each link coincide. The string-breaking phase transition is found to occur at a critical mass $m_c = 2.5$. In the small-mass regime, the ground state corresponds to a symmetric superposition of two particle-antiparticle pair configurations in which the flux-tube can pass along two distinct paths (upper-left inset) of the plaquette.}
\label{fig:vqe}
\end{figure*}
%%%%%%%%%%%%%%%%%%%%%%%%%%%%%%%%%%%%%%%%%%%%%%%%%%%%%
%%%%%%%%%%%%%%%%%%%%%%%%%%%%%%%%%%%%%%%%%%%%%%%%%%%%%

%%%%%%%%%%%%%%%%%%%%%%%%%%%%%%%%%%%%%%%%%%%%%%%%%%%%%

\section{Results and discussion}
In the following, we specialize to the $U(1)$ case of the Yang-Mills model~\eqref{eq:final-YM-hamiltonian},
and study the phenomenon of string breaking in (1+1)- and (2+1)-dimensional lattice QED which is reminiscent of \textit{confinement} in QCD. 

\subsection{Groundstate preparation}
First, we aim to represent the groundstates in different (fermionic) mass regimes using 
the gauge invariant trial states 
$\ket{\phi(\bs{\theta})} = \hat{\mathcal{U}}(\bs{\theta})\ket{\phi_\text{init}}$. Here, the initial state $\ket{\phi_\text{init}}$ is chosen as the gauge invariant bare vacuum state configuration with zero particles on the lattice sites.
The lattice configurations and open boundary conditions are assumed as sketched in Fig.~\ref{fig:vqe}$\,$a).

For our specific model we use Wilson fermions with $r=a=1$, $g = 5 - d$ and we adopt the Quantum Link Model approach~\cite{Mathis2020} for the gauge fields with a minimal truncation value of $S=0.5$ and a background electric field $\theta = 0.5\,$.
This choice encompasses either a single positive or a vanishing electric flux on each link. Note that due to our specific choice of the boundary conditions, this truncation does not affect the quality of the groundstate predictions (see Appendix~\ref{app:1d-example_model}). %configurations.
To approximate the ground state, we employ the VQE algorithm~\cite{Peruzzo2014, McClean2016} on the circuit variables $\bs \theta$, for each value of the varying mass $m$ (see Fig.~\ref{fig:vqe}).

Starting with the $d=1$ dimensional case, we define a specific family of gauge-invariant unitaries by
\begin{align} \label{eq:unitary-varform-1}
        \hat{\mathcal{U}}\big(\{\theta^k_x\}, \{\lambda_x\}\big)&=\prod_{x,k} \exp\left[i\,\Big(\hat{\psi}^\dagger_x\,A(\theta^k_x)\,\hat{U}_{(x,k)}\,\hat{\psi}_{x+k} + \text{h.c} \Big) \right] \nonumber \\ &\cdots \, \prod_{x}\exp\left[i\,\Big(\hat{\psi}^\dagger_x\,B(\lambda_x)\,\hat{\psi}_x \Big) \right]
\end{align}
which consists of hopping-like terms in order to mimic nearest-neighbor hopping dynamics, and further contains mass-like terms that essentially act as additional single-qubit post-rotations. The coupling matrices are defined as $A(\theta) = \big((0,0), (\theta,0) \big)$ and $ B(\lambda) = \big((0,0), (0,\lambda) \big)$, and $\bs{\theta} = \big(\{\theta^k_x\}, \{\lambda_x\}\big)$ denote the variational parameters. Note that the hopping terms assigned to neighboring sites commute and the products in Eq.~\eqref{eq:unitary-varform-1} are thus well-defined.

In the $d=2$ case, the variational form~\eqref{eq:unitary-varform-1} is extended by string-like terms $\hat{\mathcal{S}}(\bs{\theta})$ of the form
\begin{align}
    \hat{\mathcal{S}}\big(\{\theta^{k,l}_x\}\big) &= \prod_{x,k,l} \exp\left[i\,\Big(S_{kl} + S^\dagger_{kl}\Big) \right]\cdots \nonumber \\
    \cdots &\prod_{x,k,l} \exp\left[i\,\Big(S_{kl} + S^\dagger_{kl}\Big)\Big(S_{lk} + S^\dagger_{lk}\Big) \right],
\end{align}
where $k,l >0$ and $S_{kl} = \hat{\psi}^\dagger_x\,A\big(\theta^{k,l}_x\big)\,\hat{U}_{(x,k)}\hat{U}_{(x+k,l)}\,\hat{\psi}_{x+k+l} $ as to provide longer range explicit correlations. The need for such additional correlations is due to the groundstate configurations of the system (see left inset in Fig.~\ref{fig:vqe} c)) where the distance of the particle-antiparticle pair stretches over two links, while hopping-like terms merely produce particle-antiparticle creation on neighboring sites. On the other hand, the squared string-like terms were added to produce the corresponding groundstates that appear in a superposition.

% while in the $d=2$ case, the variational form~\eqref{eq:unitary-varform-1} is extended by string-like terms $\hat{\mathcal{S}}(\bs{\theta})$ of the form
% \begin{align}
%     \hat{\mathcal{S}}\big(\{\theta^{k,l}_x\}\big) &= \prod_{x,k,l} \exp\left[i\,\Big(S_{kl} + S^\dagger_{kl}\Big) \right]\cdots \nonumber \\
%     \cdots &\prod_{x,k,l} \exp\left[i\,\Big(S_{kl} + S^\dagger_{kl}\Big)\Big(S_{lk} + S^\dagger_{lk}\Big) \right],
% \end{align}
% where $k,l >0$ and $S_{kl} = \hat{\psi}^\dagger_x\,A\big(\theta^{k,l}_x\big)\,\hat{U}_{(x,k)}\hat{U}_{(x+k,l)}\,\hat{\psi}_{x+k+l} $ as to provide longer range explicit correlations. The coupling matrices are defined as $A(\theta) = \big((0,0), (\theta,0) \big)$ and $ B(\lambda) = \big((0,0), (0,\lambda) \big)$.

Finally, the variational operator $\hat{\mathcal{U}}(\bs{\theta})$ is translated into a quantum circuit by means of a Trotter approximation~\cite{Trotter1959, Suzuki1976} with one single Trotter step, and by using the Jordan-Wigner fermion-to-qubit mapping and a logarithmic encoding of the gauge field operators~\cite{Mathis2020}.
%Note that writing the variational form $\hat{\mathcal{U}}(\bs{\theta})$ in a product structure $\prod_k \exp{\big(i\,\hat{F}_k(\bs{\theta})\big)}$ rather than having a sum in the exponential $\exp{\big(\sum_k i\,\hat{F}_k(\bs{\theta})\big)}$ is more convenient since, although both expressions are gauge-invariant, an implementation of these exponentials will require a Trotter approximation which might not preserve the exact gauge invariance, and such an approximation error is reduced in the former expression.

The results of the variational simulation~\cite{Qiskit} are shown in Fig.~\ref{fig:vqe}$\,$ b) and c) for $d=1$ and $d=2$ dimensions, respectively. The best run (i.e., the one with the lowest groundstate energy $E_0$) out of five independent VQE optimization trials (each with randomly chosen initial parameters $\bs{\theta}$) is plotted. In fact, due to the small difference between the groundstate and the first excited state in the $d=2$ case (as explained below), more optimization trials in the VQE algorithm were needed to obtain the true groundstate.
We identify a change of behavior from a particle-antiparticle state to a single flux-string state by plotting
the particle number $N = \langle\hat{N}\rangle=\sum_x \langle \hat{\bar{\psi}}_x\hat{\psi}_x + \mathbb{I}\rangle$ as a function of the mass parameter $m$. In $d=2$ dimensions, we further plot the expectation value of the electric flux $L = \langle \hat{E}_{(x,k)}+\,\theta \rangle $ for each link in Fig.~\ref{fig:vqe} c).
In this last case, for small $m$,
the groundstate is a superposition of states where the flux is traversing the plaquette
through two possible paths (see inset in Fig.~\ref{fig:vqe} c). The degeneracy between the symmetric and antisymmetric  combination of the two paths is broken by the %less dominant 
presence of the small (as $g^2\gg 1/g^2 $) but sizable plaquette term in the Hamiltonian (see Eq.~\eqref{eq:final-YM-hamiltonian}).

We also perform, as a proof-of-principle, calculations on hardware for the $d=1$ model using the $ibmq\_ vigo$ device provided by IBM Quantum (see Fig.~\ref{fig:vqe} b). The resulting circuit~\eqref{circ:hardform} features 5 qubits and 16 cnot gates with 3 variational parameters, see Appendix~\ref{app:1d-example_hardware}.
Crucially, this strategy allows us to retrieve a qualitatively better description of the problem using much fewer parameters and entangling gates compared to heuristic circuit ans\"{a}tze (see Appendix~\ref{app:performance-vqe}).
 
%%%%%%%%%%%%%%%%%%%%%%%%%%%%%%%%%%%%%%%%%%%%%%%%%%%%%
%%%%%%%%%%%%%%%%%%%%%%%%%%%%%%%%%%%%%%%%%%%%%%%%%%%%%
%%%% FIGURE 2 REAL TIME EVOLUTION
\begin{figure}[t]
\hspace*{-0.4cm}
\includegraphics[width=0.5\textwidth]{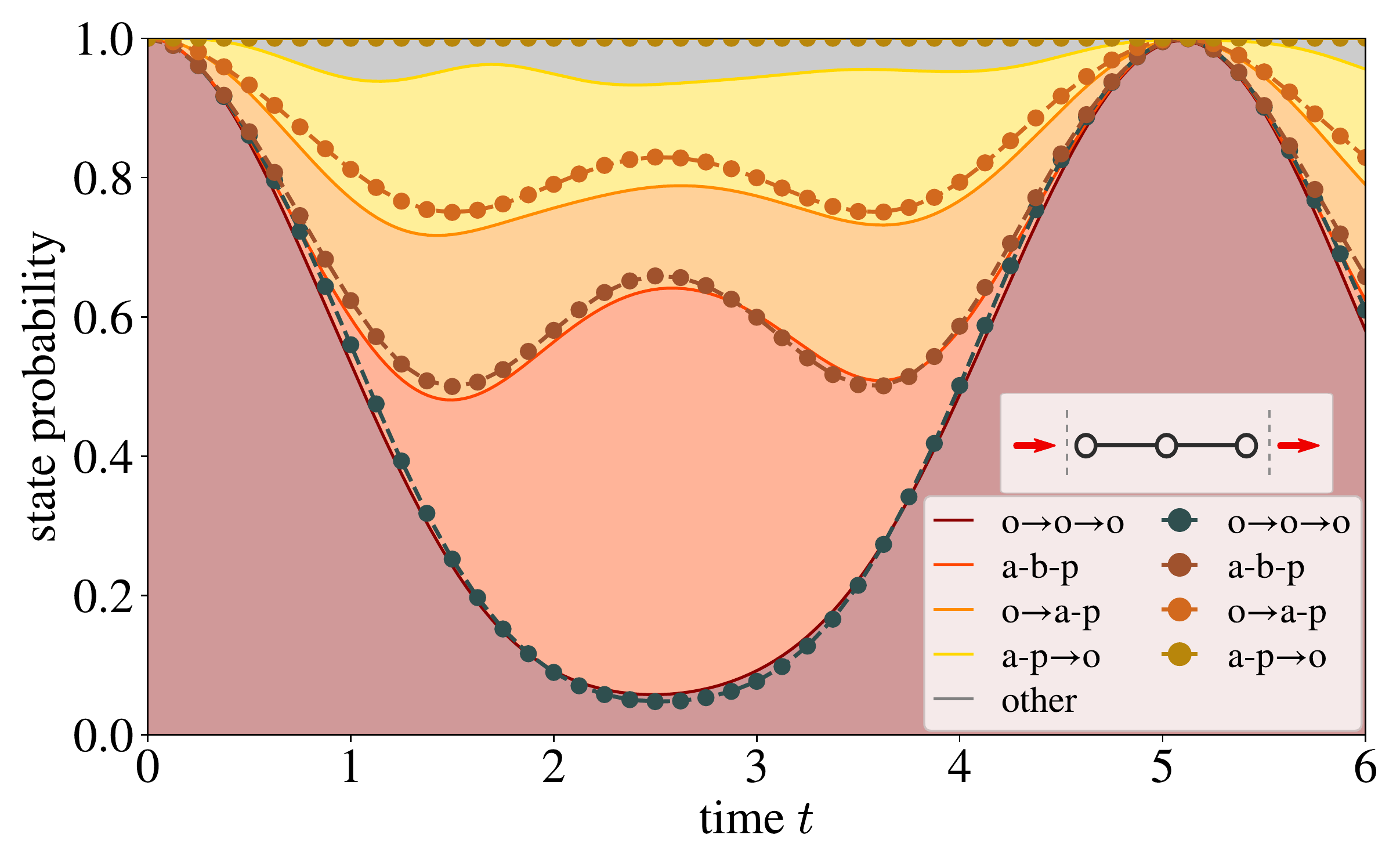}
\caption{State representability of the variational trial states in the real-time dynamics of a one-dimensional lattice with three sites and open boundary conditions (as sketched in the inset). The chosen model parameters read $r=a=1$, $g=3$, $m=1$, $S= \theta = 0.5$. The probabilities  of the different state configurations are shown as a function of time $t$ for exact evolution (solid lines) and for the optimized trial states (dotted lines). Symbols: `o': empty site, `a': anti-particle, `p': particle; `b': particle/anti-particle pair, `-': no flux line, `$\rightarrow$': flux line to the right.}
\label{fig:real-time}
\end{figure}
%%%%%%%%%%%%%%%%%%%%%%%%%%%%%%%%%%%%%%%%%%%%%%%%%%%%%
%%%%%%%%%%%%%%%%%%%%%%%%%%%%%%%%%%%%%%%%%%%%%%%%%%%%%

\subsection{Real-time propagation}
Finally, we investigate the ability of the proposed gauge-invariant quantum circuits to prepare states arising in real-time dynamics, which we will refer to as \emph{state representability}. 

We use the proposed ansatz to study the dynamics of the one-dimensional lattice Hamiltonian with three sites in the small mass regime $m\ll m_c$. The model parameters are the same as for the simulations shown in Fig.~\ref{fig:vqe} b) with explicit mass values $m=1$ and $m_c = 3.5$.
The trial states are constructed using the unitary~\eqref{eq:unitary-varform-1}. At each time step, we numerically maximize the overlap $|\langle\phi(\bs{\theta})|\psi(t)\rangle|^2 $ between the variational trial states $\ket{\phi(\bs{\theta})} $ and the exact time-evolved state $\ket{\psi(t)}$ obtained by using the matrix representation of the time propagator. 
In Fig.~\ref{fig:real-time}, we show how the trial states can capture the correct particle number oscillation characteristic of string-breaking even though the state space spanned by the trial states does not cover the entire physical Hilbert space $\mathcal{H}^\text{phys} $.

This result is of particular relevance in view of the possible use of the proposed circuits in variational time evolution algorithms~\cite{Yuan2019}.
In fact, the circuit used in this example merely requires 32 cnot gates (a value that in the variation algorithm remains constant for every time $t$) and 5 variational parameters.
On the other hand, the standard Trotterization approach  requires a much larger number of cnot gates, as a single Trotter step would feature 216 cnots, while at least $\sim 10$ Trotter steps would be needed to reach an accuracy~\cite{Mathis2020} comparable to the one reported in Fig.~\ref{fig:real-time}. In such a low mass regime, it was indeed expected from an energetic point of view that particle-antiparticle pairs would mainly form within short distances (i.e., in neighboring lattice sites) and thus, the states formed via the unitary~\eqref{eq:unitary-varform-1} would well approximate the involved states in the evolution. By going to a different (critical or large) mass regime however, one would need to adapt the trial states accordingly by extending the unitary~\eqref{eq:unitary-varform-1} with string-like terms, for instance, in order to account for a better approximation of the trial states, while coming at the expense of a larger amount of gates in order to build up the unitary. 

Future work will necessarily include an analysis of various possible combinations of the single components in the variational form toolbox in order to find the optimal trade-off between large circuits implementing the unitary and an effective sampling of the whole physical Hilbert space.

\section{Conclusions}

We introduced a general methodology to construct parametrized families of quantum circuits that preserve the gauge symmetry for $U(1)$ and Yang-Mills theories.
When applied to ground-state search algorithms, the method allows for reliable variational calculations as the trial states are always constrained within the physical submanifold of the full Hilbert space, reducing therefore the number of variational parameters.
The main advantages of realizing the Gauss law constraint at the circuit level, rather than as an energy penalty term, include:
\emph{(i)} the smoothness of the resulting energy landscape, which allows for faster optimizations,
and \emph{(ii)} 
a substantial decrease of the energy estimator variance~\cite{Bravyi2017, Kandala2017} due to the absence of the (usually large) energy penalty term.
In this work, we demonstrated the accuracy of the method in QED models featuring particles and fields as explicit degrees of freedom.

Going beyond ground-state simulations, we show that the proposed gauge invariant trial states can further efficiently represent quantum states originating from real-time evolution, 
realizing a significant reduction in circuit depth (gate count) compared to standard approaches based on the Trotterization of the time-evolution operator. We therefore anticipate the usage of the present quantum circuits in combination with variational real-time evolution algorithms~\cite{Yuan2019} to reduce the computational costs associated with the simulation of the dynamics of $SU(N)$ gauge theories, such as quantum chromodynamics. %, on quantum computers.

\section*{Acknowledgements}
The authors thank Erez Zohar for his valuable comments.

\appendix

\section{Performance of heuristic variational forms for $U(1)$ and $SU(2)$ lattice models in 1D } \label{app:performance-vqe}

In quantum field theories, the VQE~\cite{Peruzzo2014, McClean2016} can be exploited to study various groundstate properties for varying model parameters. For example, one might investigate phase diagrams of Abelian and non-Abelian gauge theories~\cite{Silvi2017,Silvi2019,Felser2020} in the regime of a finite fermion density, i.e. at an unbalance between fermionic matter and antimatter, which remain inaccessible to standard Monte Carlo simulation methods due to the notorious sign-problem.

\begin{figure}[htp]
    \centering
    \includegraphics[width=0.45\textwidth]{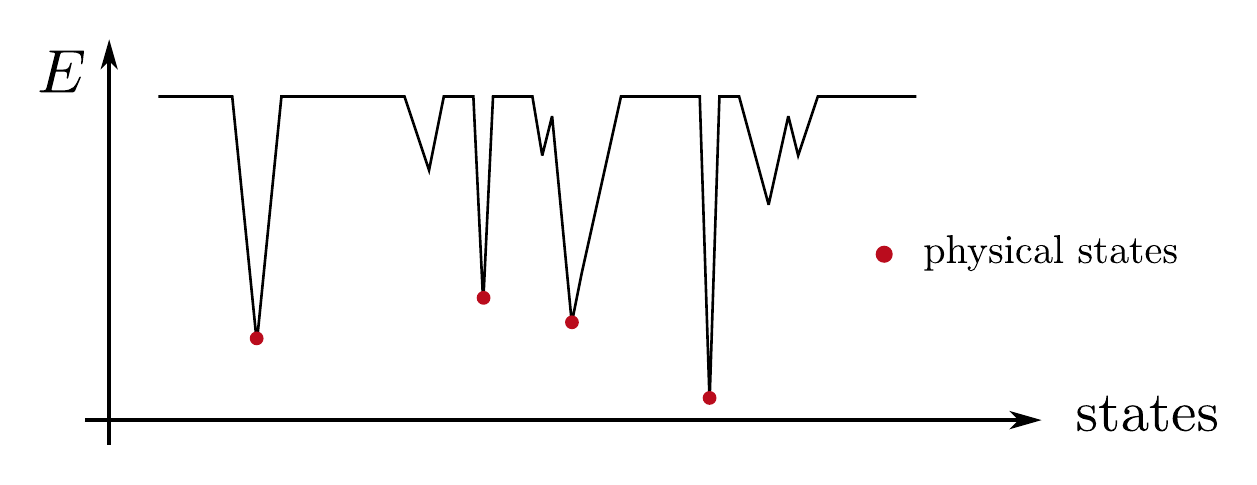}
	\caption{Graphical illustration of the corrugated energy landscape. The gauge penalty term $\hat{H}_\text{gauge}$ lifts the energy of the unphysical states in order to obtain the correct low-energy spectrum of the theory. The physical states are marked by red dots.}
	\label{fig:corrugated-landscape}
\end{figure}

When applying the VQE algorithm to a general Yang-Mills lattice gauge theory, a challenge arises in the search for an efficient ansatz for the variational form~\cite{Simon2020}. The reason for this challenge stems from the Gauss law constraint that determines the physical Hilbert subspace, $\mathcal{H}^\text{phys}\subset \mathcal{H}$ which makes up an exponentially small fraction of the total Hilbert space $\mathcal{H}$. In our implementation~ (that follows the framework of Ref.~\cite{Mathis2020}), the Gauss law constraint is imposed by adding a gauge penalty term $\hat{H}_\text{gauge}$ which raises the energy of the unphysical gauge-variant states. As a consequence, the gauge penalty term leads to a corrugated energy landscape~\cite{Berthier_2009} (Fig.~\ref{fig:corrugated-landscape}) whose global minimum is in general hard to find for classical optimization algorithms, as was observed in~\cite{Simon2020}.

In order to quantitatively confirm this hypothesis, we tested the performance of the so-called \textit{RY} and \textit{RYRZ variational form} on a small lattice QED example in $d=1$ dimensions for a varying mass parameter $m$, which is mapped to a 5-qubit system. This example is introduced in detail in the next section.

\begin{figure*}[htp]
    \hspace*{-0.7cm}
    \includegraphics[width=1.\textwidth]{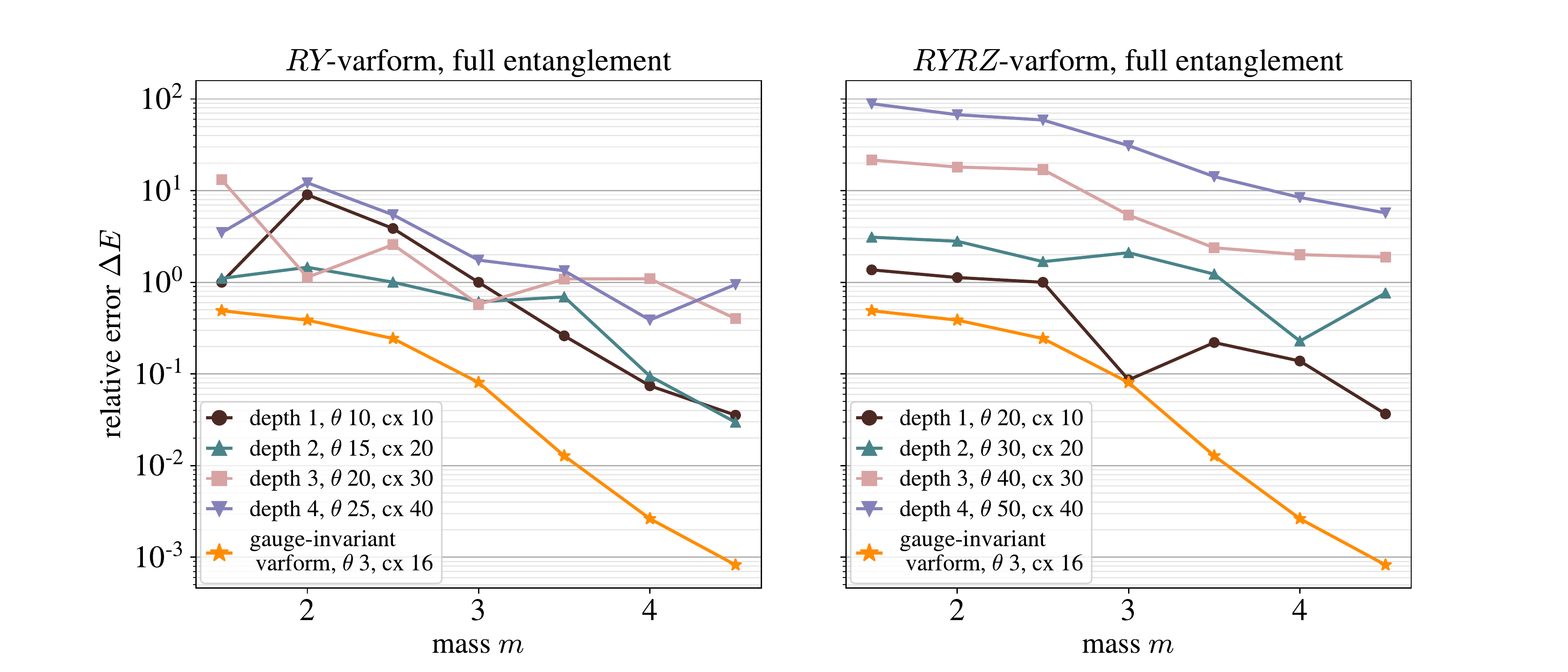}
	\caption{Performance of the $RY$ (left) and $RYRZ$ (right) variational forms with full entanglement in a VQE. The relative error $\Delta E$ in energy is plotted for varying mass parameter $m$. Each data point corresponds to a separate VQE simulation. The "$\theta$" denotes the number of parameters in the variational form while "$cx$" stands for the count of $CNOT$ gates in the corresponding circuit.}
	\label{fig:ry-ryrz-performance}
\end{figure*}

The $RY$ and $RYRZ$ are two standard heuristic variational forms $\hat{\mathcal{U}}(\bs{\theta})$ that are constructed such that a wide range of states in a general Hilbert space can be covered without taking into account the underlying structure of the corresponding physical system. To do so, these variational forms systematically entangle all of the qubits and apply parametrized rotations $R_y$ for $RY$ or $R_y$ and $R_z$ for $RYRZ$, respectively, on each single qubit. 

The performance of these two variational forms is shown in Fig.~\ref{fig:ry-ryrz-performance}. The relative energy difference $\Delta E = |E_\text{vqe} - E_0|\,/\,|E_0|$ is plotted as a function of varying mass parameter $m$ and for increasing depth of the $RY$ and $RYRZ$ variational forms. Each data point in the plot corresponds to a single VQE run (in statevector simulation) which achieved the lowest energy estimation $E_\text{vqe}$ out of a total of 10 independent optimization runs (each starting from different randomly chosen initial parameters) for each mass value. We used the \textit{Cobyla}  optimization algorithm for the classical optimization procedure. The initial parameters $\bs{\theta}^{(0)}$ were randomly chosen from a uniform distribution in $[-\pi, \pi]$ and the initial state $\ket{\phi_\text{init}}$ was given by the bare vacuum state, i.e. the unique gauge-invariant state with zero particles on each lattice site. For comparison, we further included the performance of the gauge-invariant variational form~\eqref{eq:unitary-varform-1} with $A(\theta) = \big((0,0), (\theta,0) \big)$ and $ B(\lambda) = \big((0,0), (0,\lambda) \big)$. Consequently, the matrices $A$ couple only the second spinor component at a site $x$ to the first spinor component at an adjacent site $x+k$. This implies that hopping terms assigned to neighboring sites commute and thus, the products in Eq.~\eqref{eq:unitary-varform-1} are well-defined. Furthermore, for each site that sits in a corner of the finite lattice, there is always one spinor component that is left untouched by this variational form, allowing to effectively remove the qubits that encode these spinor components~\cite{Simon2020}. The corresponding quantum circuit for a one-dimensional lattice system with two sites is shown in~\eqref{circ:hardform}.

\begin{figure*}[htp]
    \hspace*{-0.75cm}
    \includegraphics[width=1.\textwidth]{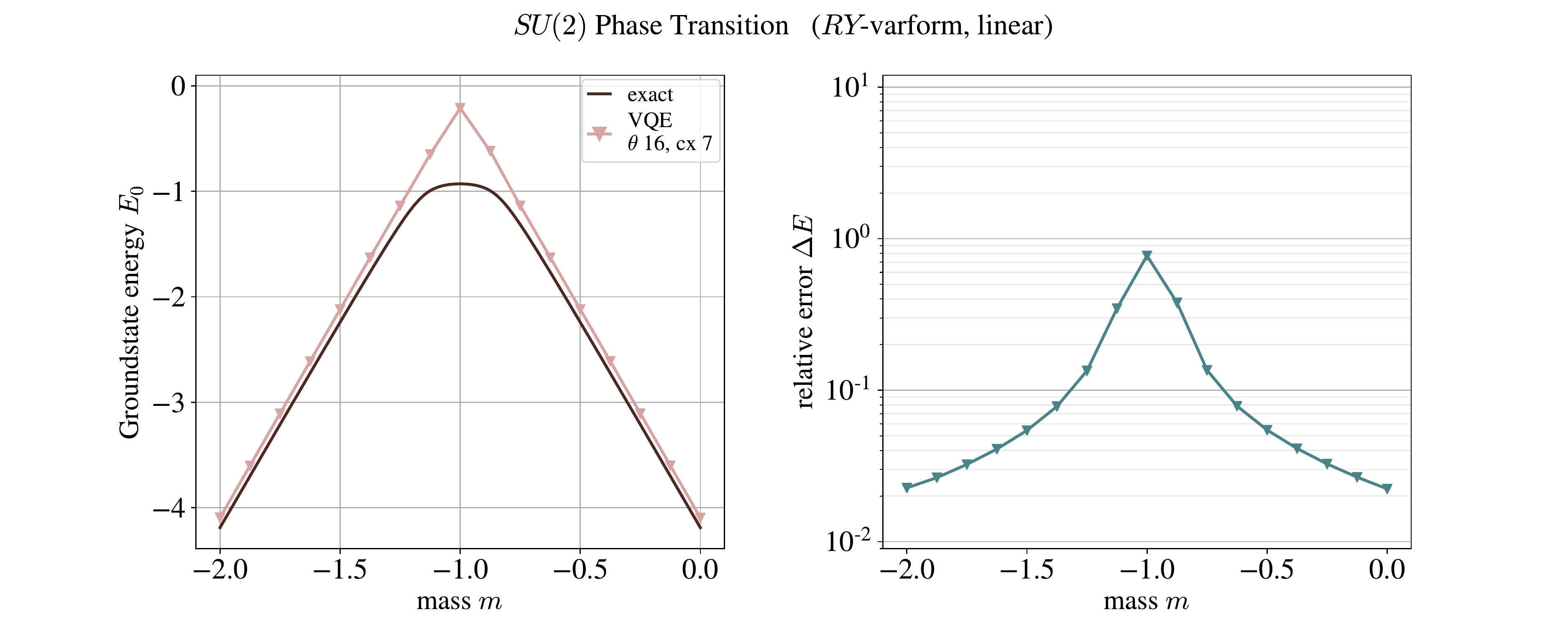}
	\caption{Results for the one-dimensional integrated $SU(2)$ model with two lattice sites and performance of the $RY$ variational form using linear entanglement with 16 parameters and 7 $CNOT$ gates. The groundstate energy $E_0$ (left) and the relative energy difference $\Delta E$ (right) are shown as a function of mass $m$.}
	\label{fig:ry-su2-performance}
\end{figure*}

From the results in Fig.~\ref{fig:ry-ryrz-performance} it is visible that the standard heuristic variational forms struggle to find the correct groundstate energy, resulting in a poor approximation of $E_0$. Raising the circuit depth merely aggravates the performance in spite of a higher number of degrees of freedom. The origin for such large errors lies in the large regularizing coefficient $\lambda = 20$ in the gauge penalty term $\hat{H}_\text{gauge}$ that leads to high energy contributions in the order of $\sim 10^2$ but which is needed to fix the correct low-lying gauge-invariant energy spectrum. Furthermore, it was observed that the performance of the standard variational forms greatly depended on the initial set of parameters $\bs{\theta}^{(0)}$. Therefore, these standard variational form ans\"{a}tze are not promising for larger system sizes and non-abelian gauge theories. 

In contrast, the gauge-invariant variational form clearly outperforms the standard heuristic ans\"{a}tze by one or two orders of magnitude. At the same time, it requires much less parameter degrees of freedom which is beneficial for the classical optimization procedure, even for larger sizes of the physical system. 

We further tested the performance of the standard heuristic $RY$ variational form on a $d=1$ dimensional $SU(2)$ gauge theory model where the Gauss law constraint is incorporated in the Hamiltonian and the gauge degrees of freedom are effectively erased~\cite{Sala2018}. Since the Hilbert space of this model only contains gauge invariant states, we expect the $RY$ variational form to achieve better convergence to the groundstate energy.

The integrated $SU(2)$ model was implemented on $M=2$ lattice sites with lattice spacing $a=1$, Wilson parameter $r=1$, coupling constant $g=3$ and boundary conditions $\hat{R}^\alpha_0 = 0$, $\forall\,\alpha$. The groundstate energy $E_0$ was examined as a function of varying (negative) mass parameter $m$. In the regime where $m\approx -r/a$, the model exhibits a specific type of behavior reminiscent of a phase transition from a bare vacuum phase with no particles to a charge-crystal phase where all lattice sites are fully occupied~\cite{Felser2020}. Such a phase transition occurs when the prefactor $\big(m+r/a\big)$ of the mass term $\sum_n \hat{\bar{\psi}}_n\hat{\psi}_n $ in the system Hamiltonian changes its sign from positive to negative since in the former case, the production of particles costs energy while in the latter case, it becomes energetically favorable to create particles pairs.

For various mass values $m$, we ran a separate ideal VQE for each $m$ using the $RY$ variational form with linear entanglement, i.e. where the qubits are not all pairwise entangled but only neighboring qubits are entangled, leading to a reduced number of $CNOT$ gates. We used the \textit{Cobyla} optimization algorithm for the classical optimization procedure. The exact groundstate energy curve was calculated by exact diagonalization. The results are plotted in Fig.~\ref{fig:ry-su2-performance}.

From these results, it is visible that the $RY$ variational form shows a better performance with smaller relative errors $\Delta E  = |E_\text{vqe} - E_0|\,/\,|E_0|$ for the integrated $SU(2)$ model than for the general scalable QED model with a gauge penalty term. However, the error increases in the regime of the phase transition, indicating that this heuristic variational form is not able to capture the essential properties of the groundstate in the region of the phase transition where the quantum fluctuations, arising from the hopping term in the system Hamiltonian, start to become significantly large.

It has further been checked that changing the linear entanglement to a full entanglement or raising the depth of the variational form did not lead to significant improvements for the overall performance.

To conclude, this short analysis corroborates the need of physically motivated and structured variational forms which respect the symmetries of the Yang-Mills theory at hand. Furthermore, note that heuristic standard variational forms are usually constructed at the qubit level and thus will depend on the exact encoding scheme for the gauge and matter field degrees of freedom.

\section{Explicit lattice QED example in 1D on real quantum hardware}

Here we present a detailed description of the (1+1)-dimensional lattice QED system exhibiting the phenomenon of string breaking, which was simulated on a few-qubit superconducting quantum device.

\subsection{The model} \label{app:1d-example_model}
In the following, we consider a one dimensional lattice with $M=2$ sites and one link with open boundary conditions. It is convenient to relabel the matter fields by $\hat{\psi}_x \to \hat{\psi}_n$, the flux operator by $\hat{L}_{(x,k)}\to\hat{L}_n$ and the link variable by $\hat{U}_{(x,k)}\to \hat{U}_n$. Recall that the QED Hamiltonian~\eqref{eq:final-YM-hamiltonian} for dimension $d=1$ including the gauge penalty term reads
\begin{equation}\label{eq:1d-qed-hamiltonian}
    \begin{split}
    \hat{H}_\text{QED} = &\sum_{n=1}^{M-1}  \frac{1}{2a} \left( \hat{\bar{\psi}}_n \left[ i\gamma^1 + r\right]\hat{U}_n\hat{\psi}_{n+1} + \text{h.c.} \right) \\
    &+ \sum_{n=1}^{M} \left( m + \frac{r}{a} \right)\hat{\bar{\psi}}_n \hat{\psi}_n  + \frac{g^2a}{2} \sum_{n=1}^{M-1} \hat{L}^2_n \\
    &+ \lambda \sum_{n=1}^M \hat{G}^2_n\\
    \end{split}
\end{equation}
with $\hat{G}_n = \hat{L}_n - \hat{L}_{n-1} - \hat{\psi}^\dagger_n \hat{\psi}_n\,, $ and where $\hat{L}_0$ and $\hat{L}_M$ are not dynamical variables but are fixed by the choice of the open boundary conditions.

In addition, we allow for a non-trivial constant background electric field $\theta$ to be added to the model~\cite{Funcke2020}, which is taken into account by shifting the electric flux field operator by a constant amount,
\begin{equation}
    \hat{L}_n \ \rightarrow \ \hat{L}_n + \theta \quad \Rightarrow \quad  \hat{H}_\text{elec} \ \rightarrow \ \frac{g^2a}{2} \sum_{n=1}^{M-1} \left(\hat{L}_n + \theta \right)^2
\end{equation}
where $\,\theta \equiv \theta \cdot \mathbb{I}\,$ acts as an identity operator times a real constant. Therefore, such a background electric field effectively shifts the allowed electric flux eigenvalues $m_s$ by an amount of $\theta$. For example, for a spin value $S=1.5$ in the quantum link model formulation of the Abelian gauge fields and a background electric field $\theta = 0.5$, the allowed flux eigenvalues are shifted by $m_s \in \{-1.5,-0.5,0.5,1.5\}\,\to\,m_s \in \{-1,0,1,2\}$. As a result, the flux eigenvalues are not centered around the zero flux value anymore, thereby breaking the spatial symmetry of the physical system. Nevertheless, for certain choices of the boundary conditions or for high enough spin values $S$, the low-lying energy spectrum will not be affected by this shift as will be the case in the model that we are considering. On the other hand, the background electric field will allow us to reduce the number qubits needed to store the relevant states of the model.

\begin{figure}[ht]
    \centering
    \includegraphics[width=0.3\textwidth]{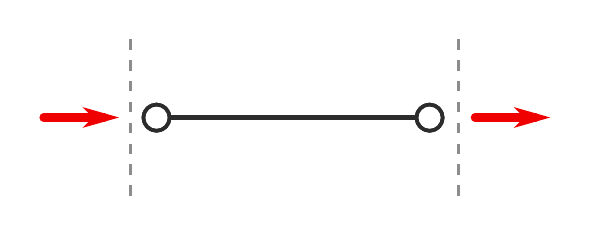}
	\caption{One dimensional lattice model with two sites and one link. The red arrows denote the flux boundary conditions at both ends of the lattice. The boundary conditions fix the allowed physical states by means of the Gauss law constraint.}
	\label{fig:1d-qed-obc}
\end{figure}

In our simulations, we choose the model parameters $g=4,\ a=1,\ r=1,\ \lambda=20$, the spin value $S=0.5$ with a background electric field $\theta = 0.5$, and boundary conditions as depicted in Fig.~\ref{fig:1d-qed-obc}.

A total of 5 qubits are needed to represent the total Hilbert space of this lattice system with spin value $S=0.5$, namely two qubits per site and one qubit for the link in between, thus spanning a Hilbert space $\mathcal{H}$ of dimension $2^5 = 32$. Furthermore, the background electric field $\theta= 0.5$ shifts the two allowed flux eigenvalues by $m_s\in\{-0.5,0.5\} \,\to\, m_s\in\{0,1\}$.

\begin{figure}[htp]
    \centering
    \vspace*{-0.3cm}
    \includegraphics[width=0.42\textwidth]{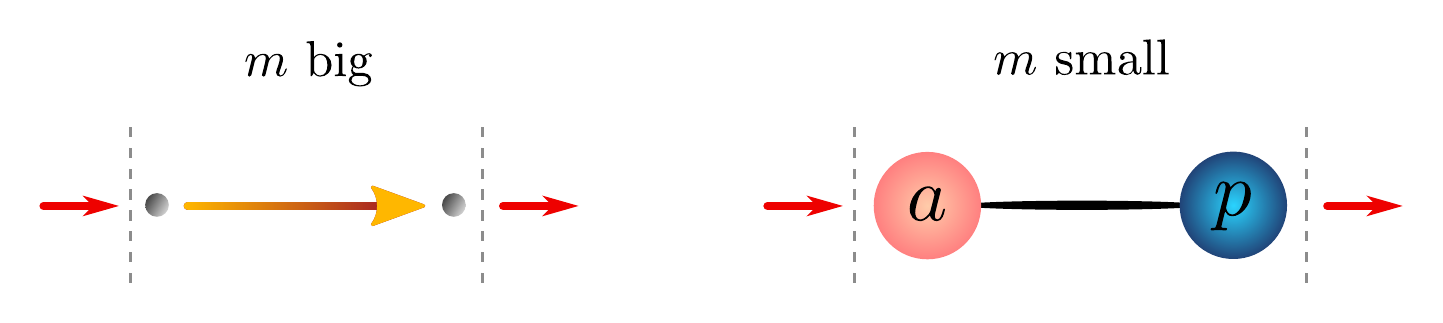}
	\caption{The groundstate configurations depending on the mass parameter $m$. When $m\gg g$ is big, then the flux string contains not enough energy in order to break (left). In the opposite case when $m\ll g$ is small, it is energetically more favorable to break the string by creating a particle-antiparticle pair (right).}
	\label{fig:1d-qed-gs}
\end{figure}

\begin{figure}[htp]
    \centering
    \vspace*{-0.3cm}
    \includegraphics[width=0.4\textwidth]{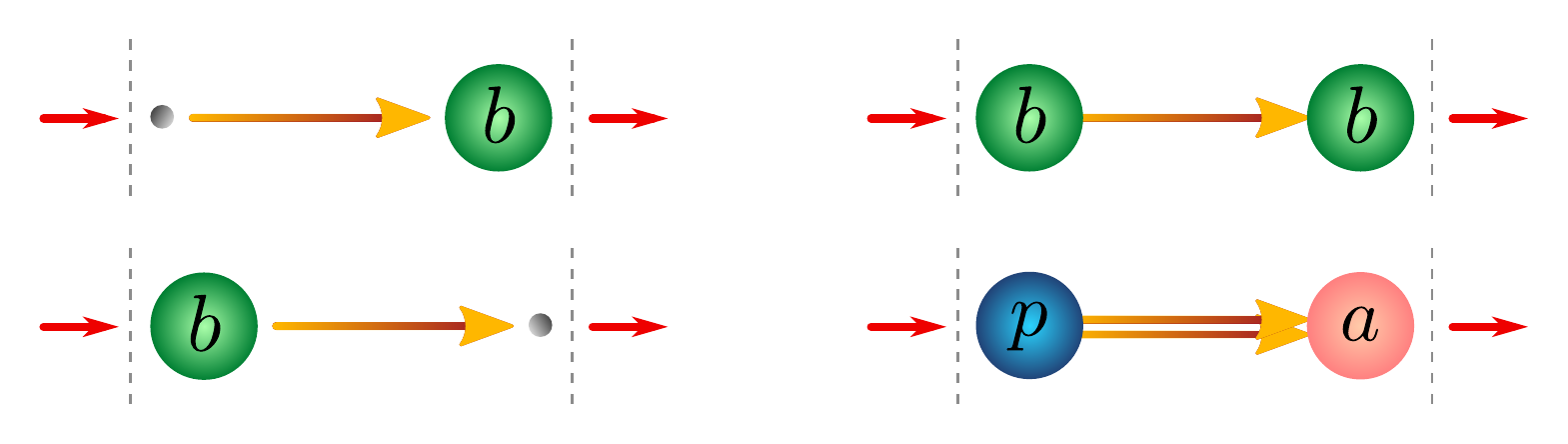}
	\caption{Complete set of gauge invariant state configurations in addition to the two groundstates shown in Fig.~\ref{fig:1d-qed-gs}. The state with a particle-antiparticle pair is not contained in the truncated state description for $S=0.5$ with $\theta = 0.5$.}
	\label{fig:1d-qed-gaugeinv}
\end{figure}

Note that out of these 32 states, only 5 states corresponds to physical states. The possible gauge invariant state configurations that are allowed by the Gauss law constraint with this choice of boundary conditions are shown in Fig.~\ref{fig:1d-qed-gs} and in Fig.~\ref{fig:1d-qed-gaugeinv}. 

From these configurations it is clearly visible that our choice of the small spin value $S=0.5$ with $\theta=0.5$ does not affect the groundstate configurations of the model~\eqref{eq:1d-qed-hamiltonian}: First, all gauge invariant states contain only fluxes in positive direction which is consistent with having possible flux values $m_s\in\{0,1\}$. Second, the only physical state that is not representable with this choice of flux eigenvalues is the one with a particle-antiparticle pair connected by two flux strings. Nevertheless, such a state will always correspond to a higher energy state configuration than the two states in Fig.~\ref{fig:1d-qed-gs} and therefore, will not occur as a groundstate of this model. As a conclusion, choosing the values $S=0.5,\  \theta=0.5$ instead of $S=1,\  \theta=0$ allowed us to effectively remove one qubit in order to represent the gauge fields on the link.

\begin{figure*}[htp]
    \centering
    \vspace*{-0.38cm}
    \hspace*{-0.95cm}
    \includegraphics[width=1.1\textwidth]{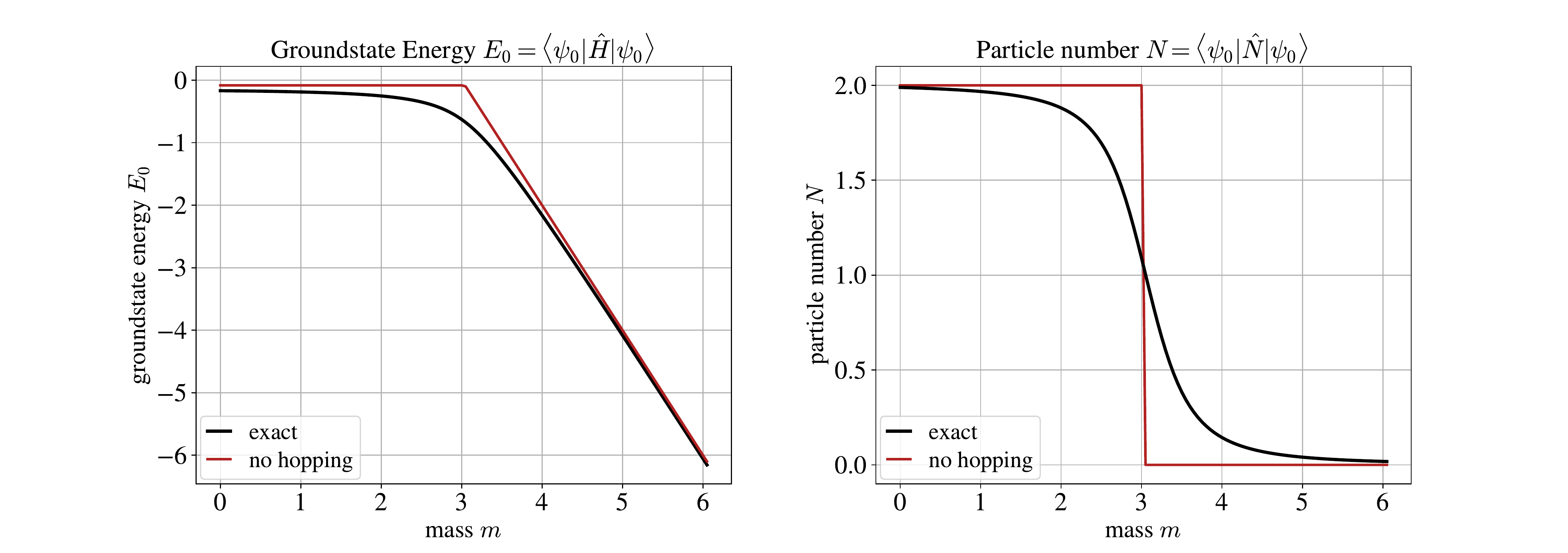}
	\caption{The groundstate energy $E_0 = \bra{\psi_0} \hat{H}\ket{\psi_0}$ (left) and the corresponding particle number $N = \bra{\psi_0} \hat{N}\ket{\psi_0}$ (right) as a function of mass $m$. The chosen model parameters in~\eqref{eq:1d-qed-hamiltonian} are $g=4,\ a=1,\ r=1,\ \lambda=20$, $S=0.5$, $\theta = 0.5$ resulting in a phase transition at the critical mass value $m_c = 3$. The black curves correspond to the full model~\eqref{eq:1d-qed-hamiltonian} while the red curves displays the groundstates for vanishing hopping term.}
	\label{fig:1D-qed-exact-plot}
\end{figure*}

Now let us analyze the groundstate for varying mass parameter $m$. First, it is clear that the only possible groundstates that are to be considered are those depicted in Fig.~\ref{fig:1d-qed-gs} since all other gauge invariant states shown in Fig.~\ref{fig:1d-qed-gaugeinv} will always have a higher energy independently of the mass value $m$. As a next step, assume that the Hamiltonian~\eqref{eq:1d-qed-hamiltonian} would not contain any hopping term. In that case, the Hamiltonian is already diagonal and the groundstate would depend on whether the mass term $\hat{H}_\text{mass}$ or the electric term $\hat{H}_\text{elec}$ has a smaller energy. The corresponding critical mass $m_c$ is simply calculated by equating the total energy of the two state configurations in Fig.~\ref{fig:1d-qed-gs},
\begin{equation}
    2 \cdot\left(m_c + \frac{r}{a} \right) \,=\, \frac{g^2a}{2} \quad \Rightarrow\quad m_c \,=\,  \frac{g^2a}{4} - \frac{r}{a} \ .
\end{equation}
and thus, the groundstate energy $E_0$ as a function of mass $m$ will have a discontinuity in its first derivative, see Fig.~\ref{fig:1D-qed-exact-plot}. Note that one might expect the groundstate energy for vanishing hopping term in the small mass regime $m\ll m_c$ to be equal to zero since both the electric field term and the mass vanish for the gauge invariant particle-antiparticle pair state. However, the small negative energy contribution that is visible in the left graph originates from the Wilson term which contains a hopping like term proportional to $r$ and which was not set to zero.

By adding the non-diagonal hopping term, the groundstate energy in the limiting regimes $m\gg m_c$ and $m\ll m_c$ stays approximately the same while in the critical mass regime $m\approx m_c$, strong quantum fluctuations are introduced that smooth out the groundstate energy and the particle number curves with $\hat{N}=\sum_n \big(\hat{\bar{\psi}}_n\hat{\psi}_n + \mathbb{I})$ as is shown in Fig.~\ref{fig:1D-qed-exact-plot}. As a result, a phase transition is observed which corresponds to the flux string breaking.

\begin{figure*}[htp]
    \centering
    %\hspace*{-0.36cm}
    \includegraphics[width=1.\textwidth]{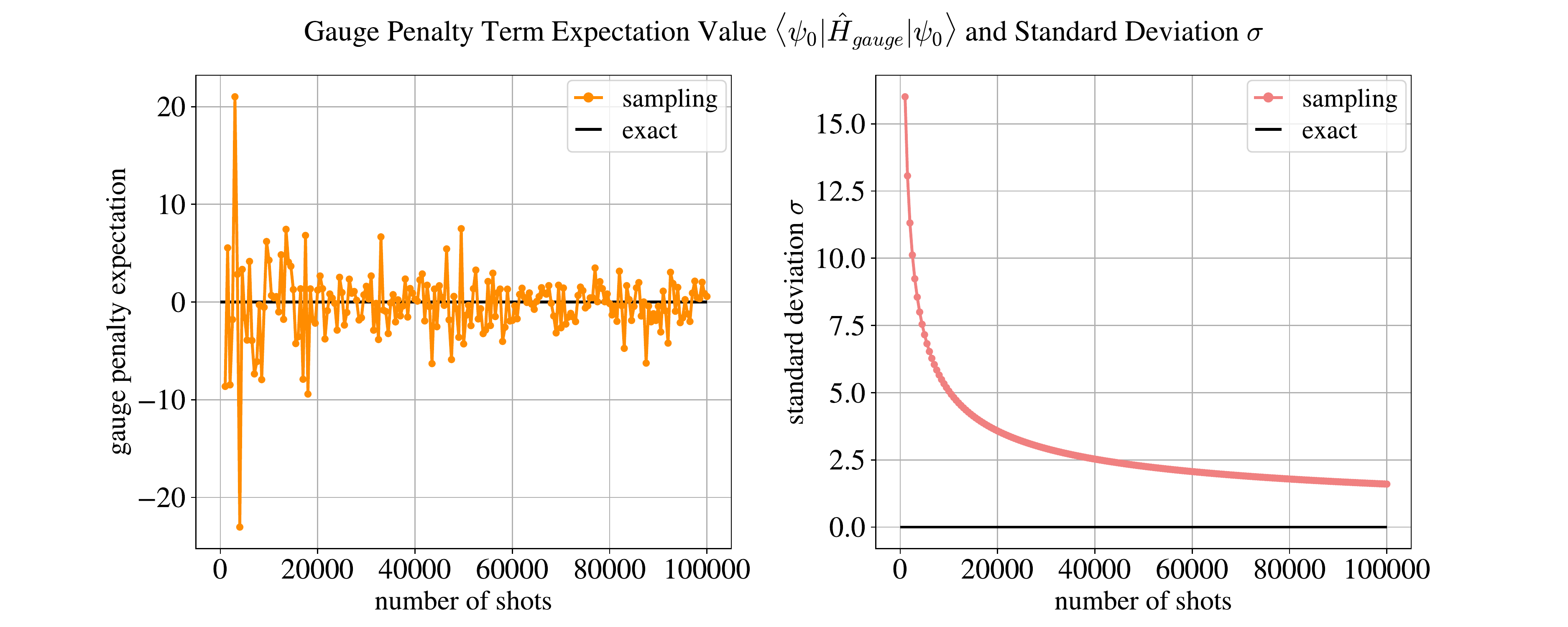}
	\caption{Statistical sampling of the gauge penalty term expectation value $\bra{\psi_0} \hat{H}_\text{gauge}\ket{\psi_0}$ (left) and the corresponding standard deviation $\sigma$ (right) with respect to a superposition of gauge invariant states. The regularizing parameter $\lambda=20$ gives rise to large statistical fluctuations around the theoretical zero expectation value.}
	\label{fig:sampling-gauge-plot}
\end{figure*}

\subsection{Statistical sampling errors due to the gauge penalty term} \label{app:1d-example_sampling}

A problem arises when evaluating the parametrized energy expectation value on a real quantum hardware since it can not be exactly calculated as in an ideal simulation, but the parametrized energy is rather statistically sampled from various measurements of the same circuit. This follows from the postulates of quantum mechanics where a measurement outcome of an observable randomly occurs with a probability determined by the wavefunction and thus, the outcome expectation value corresponds to an average of all the outcomes.

This fact can lead to problematic consequences in a VQE algorithm, especially due to the gauge penalty term $\hat{H}_\text{gauge}$ with a large regularizing coefficient $\lambda$, since the statistical sampling errors might become too big leading to large non-zero energy contribution and thus, the correct groundstate energy can not be recovered anymore, although the gauge penalty term should in principle vanish for gauge invariant states.

In our lattice gauge theory models, the sampled expectation value $\bra{\phi(\bs{\theta})}\hat{H}_\text{gauge}\ket{\phi(\bs{\theta})}$ of the gauge penalty term will in practice only approximately equal to zero for physical states. Since non-zero sampled expectation values will be proportional to the regularizing parameter $\lambda$ which is usually chosen to be much larger than the other model parameters, the energy evaluations in a VQE algorithm will be largely disturbed. 

To certify this hypothesis, we sampled the expectation value of the gauge penalty term $\lambda\,\hat{H}_\text{gauge}$ with $\lambda=20$ with respect to an equal superposition of the two gauge invariant groundstates shown in Fig.~\ref{fig:1d-qed-gs} with increasing number of shots (i.e. measurements) by using IBM's $\mathtt{Qiskit}$ \textit{QASM}-simulator~\cite{Qiskit}. Each weighted Pauli string in the gauge penalty term was measured separately. The results are plotted in Fig.~\ref{fig:sampling-gauge-plot}. It is observed that the statistical fluctuations are of the same order of magnitude as the energy scale of the groundstate energies of the considered model~\eqref{eq:1d-qed-hamiltonian}. Note that if one is interested in the groundstate only, the regularization value $\lambda$ could in principle be chosen smaller (as long as it lifts the energy of the unphysical states to an amount which is bigger than the energy of the groundstate). However, higher excited states become important in a real-time evolution in general and thus, the regularization value must increase correspondingly to ensure that the dynamics remain physical~\cite{Simon2020}.

\begin{figure}[htp]
    \centering
    \hspace*{-0.66cm}
    \includegraphics[width=0.5\textwidth]{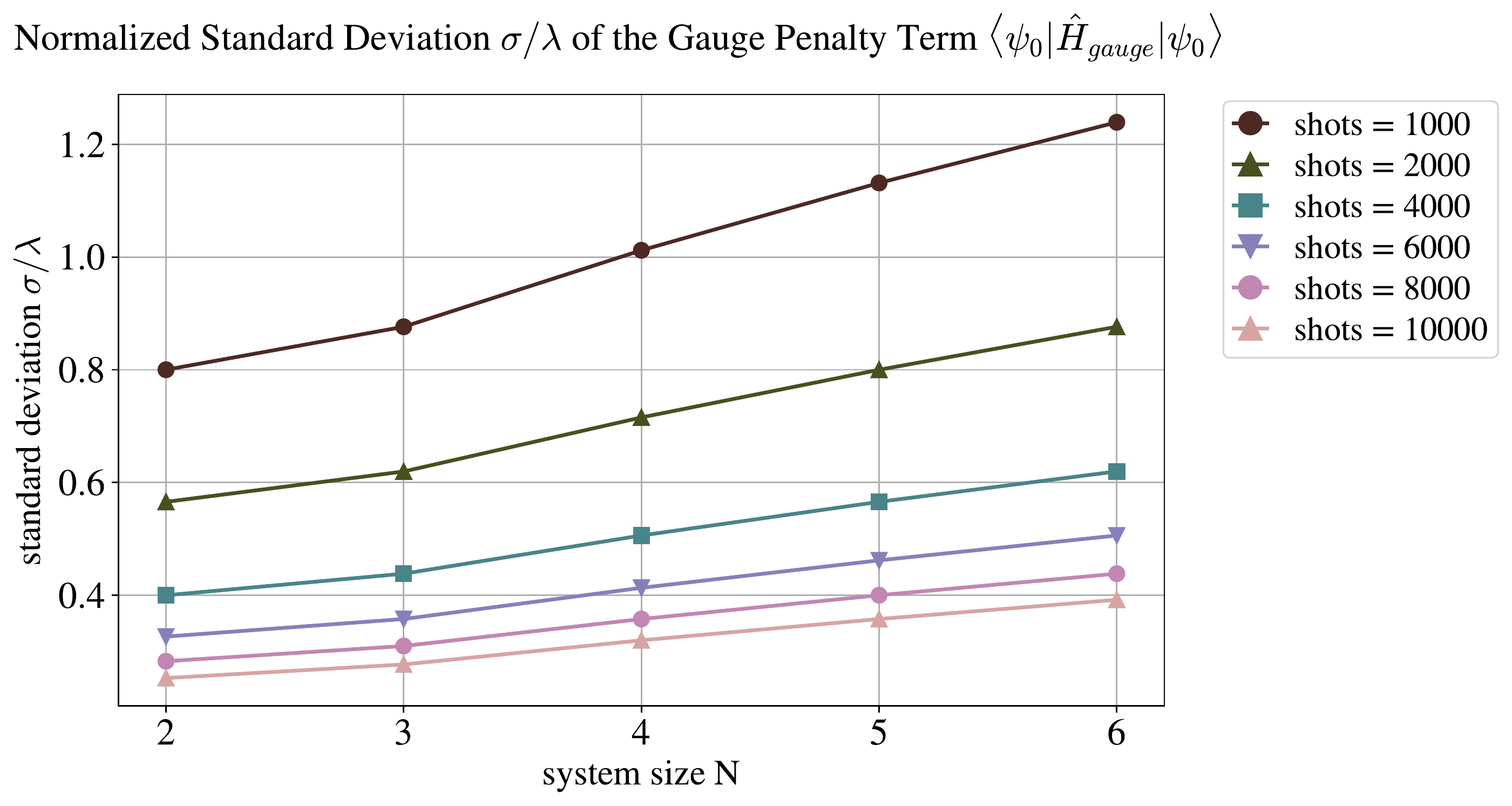}
	\caption{Normalized standard deviation $\sigma/\lambda$ of the statistical sampling error in measuring the gauge penalty term expectation value $\bra{\psi_0} \hat{H}_\text{gauge}\ket{\psi_0}$ in $d=1$ for growing system size (i.e., number of lattice sites) $N$ for different fixed number of shots. The regularization parameter is chosen as $\lambda = 20$.}
	\label{fig:sampling-gauge-plot-2}
\end{figure}

It is further shown in Fig.~\ref{fig:sampling-gauge-plot-2} how such statistical sampling errors increase for a one-dimensional lattice with increasing number of lattice sites. We observe an enhancement of the standard deviation with increasing system size, hinting that such statistical noise might become intractable for LGT models with large lattices.

The observed issue can be circumvented in different ways. First, it might happen that the weighted Pauli strings $P$ which make up the gauge penalty term $\hat{H}_\text{gauge} = \sum_P \lambda_P P$ all commute with each other, as is the case in a Jordan-Wigner fermion-to-qubit encoding of this term, for example. Then, the Pauli strings can all be simultaneously measured and therefore, the same state measurement-outcome statistics can be used to evaluate the expectation value of all the Pauli strings in a \textit{grouped} manner~\cite{McClean2016, Bravyi2017,Kandala2017}, resulting in a vanishing statistical error. However, the exact form of the Pauli strings will depend on the specific encoding of the gauge and matter fields and the Pauli strings may not pairwise commute in general. Furthermore, the exact choice of grouping the Pauli string bases might affect the overall sampling efficiency of evaluation of the total Hamiltonian.

\begin{figure*}[htp]
    \centering
    \hspace*{0.1cm}
    \includegraphics[width=0.65\textwidth]{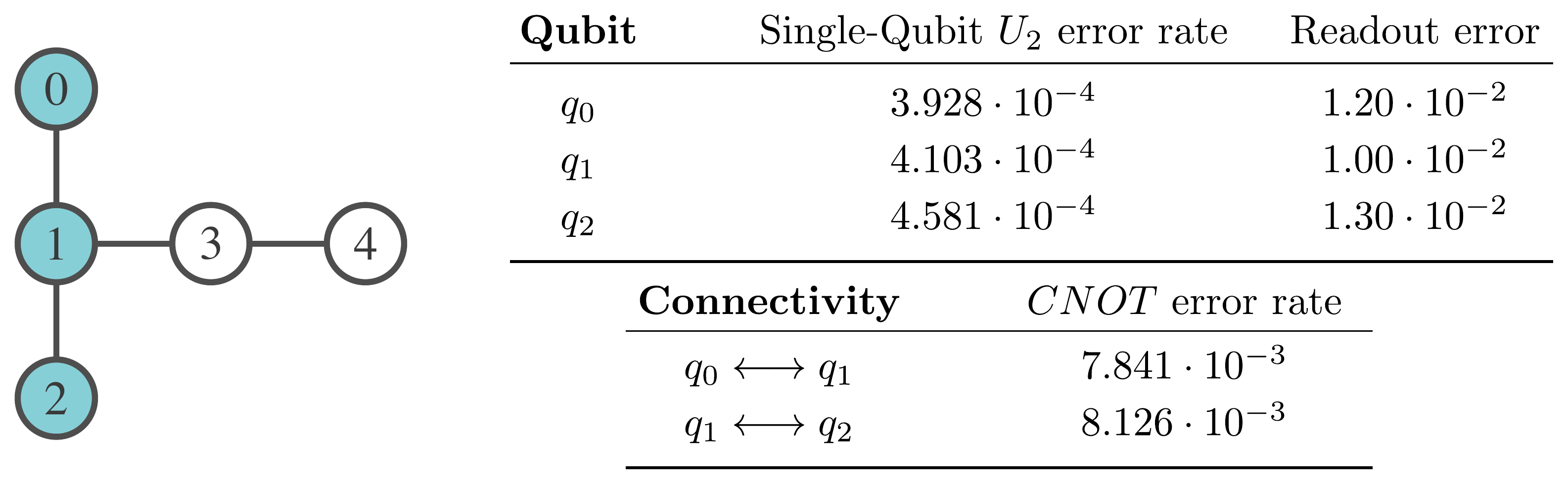}
	\caption{The \textit{ibmq\_vigo} quantum hardware with the qubit configuration topology (left) and the corresponding gate- and readout error rate specifications (right). The three qubits colored in blue correspond to the qubits used in the quantum simulation. }
	\label{fig:ibmq-vigo}
	\vspace{-0.45cm}
\end{figure*}

In our simulations we pursue a different strategy. Since we are working with a variational form that is gauge invariant by construction, there is in principle no need to additionally impose the Gauss law constraint as a penalty term at all, provided that the initial state $\ket{\phi_\text{init}}$ used in the VQE algorithm corresponds to a physical gauge invariant state configuration. As a result, the variational form samples only from the physical Hilbert space leading to the correct low-lying energy spectrum and thus, we will simply set the gauge penalty term to zero. It is observed that with vanishing gauge regularizing parameter $\lambda=0$, the statistical fluctuations remain small compared to the energy scale set by the chosen model parameters.

Nevertheless, in practice there might be a small probability that the gauge invariant variational form \textit{does} sample from a set of unphysical states depending on the form of the coupling matrices $A(\bs{\theta}),\ B(\bs{\theta})$ in~\eqref{eq:unitary-varform-1} due to Trotter approximation errors. The Trotter error in the approximation can be made arbitrarily small by increasing the number of Trotter steps at the cost of enlarging the circuit gate complexity. Note that the explicit gauge invariant variational form~\eqref{circ:hardform} that is used in our simulations consists of four Pauli strings as a result of the Jordan-Wigner mapping which all pairwise commute. Thus, there is no Trotter approximation error in our simulations.

There exists a third approach as was already pointed out in~\cite{Simon2020} that makes use of so-called Gauss law oracles~\cite{Stryker2019}. Such Gauss law oracles can be included in the VQE algorithm as a subroutine to check whether the produced variational form $\ket{\phi(\bs{\theta})}$ is actually gauge invariant or not. In that way, one can make sure that the small error produced by the Trotterization does not lead to an unphysical energy spectrum even when the gauge penalty term is set to zero.

\subsection{Quantum hardware simulation} \label{app:1d-example_hardware}

As already mentioned above (or as can be directly observed in the circuit~\eqref{circ:hardform}), the first and the last fermionic qubits "$\text{fermionic}_0$" and "$\text{fermionic}_3$" are not entangled with the other qubits since only single-qubit gates are applied to these two qubits. As a consequence, the parametrized wavefunction $\ket{\psi(\bs{\theta})}$ has a product form with respect to these two qubits which allows us to factorize the parametrized energy as
\begin{equation}
    \begin{split}
        E(\bs{\theta}) &= \bra{\phi(\bs{\theta})}\hat{H}\ket{\phi(\bs{\theta})} = \sum_{P\in\mathcal{P}_5} \lambda_P \bra{\phi(\bs{\theta})}P\ket{\phi(\bs{\theta})} \\
        &= \sum_{P\in\mathcal{P}_5} \lambda_P\, \bra{\phi_0(\bs{\theta})}P_0\ket{\phi_0(\bs{\theta})}\cdot\bra{\phi_3(\bs{\theta})}P_3\ket{\phi_3(\bs{\theta})}\\
        &\hphantom{= \sum_{P\in\mathcal{P}_5} \lambda_P}\cdot\bra{\phi_{124}(\bs{\theta})}P_{124}\ket{\phi_{124}(\bs{\theta})}
    \end{split}
\end{equation}
with $ P = P_0\otimes P_1\otimes P_2\otimes P_3\otimes P_4\,$, $\ket{\phi(\bs{\theta})}= \ket{\phi_0(\bs{\theta})}\otimes \ket{\phi_3(\bs{\theta})}\otimes \ket{\phi_{124}(\bs{\theta})}\, $, and $\hat{H}=\sum_{P\in\mathcal{P}_5} \lambda_P\,P\,,\quad \lambda_P \in\mathbb{R}$. The Pauli strings $P\in \mathcal{P}_5$ that act on a 5-qubit register with $P_{124} \equiv P_1\otimes P_2\otimes P_4$ and where $\ket{\phi_{124}(\bs{\theta})}$ denotes the wavefunction corresponding to the qubits with index 1, 2 and 4. The single-qubit terms of the form $\bra{\phi_i(\bs{\theta})}P_i\ket{\phi_i(\bs{\theta})}$ can be efficiently calculated on a classical computer since they basically correspond to $2\times 2$ matrix multiplications. As a conclusion, only the non-trivial entangled part of the wavefunction $\ket{\phi_{124}(\bs{\theta})}$ will have to be evaluated on the quantum device.

This observation will be exploited in our quantum simulation in order to reduce the required qubit resources from 5 to 3 qubits, thereby effectively lowering noise errors that are present in a simulation with real quantum hardware.

With the preliminary work in the previous paragraphs, we consequently ran a separate VQE algorithm in order to search the groundstate of the model for each distinct mass value $m$ by using the specific variational form~\eqref{eq:unitary-varform-1} with three parameters, and whose implementing circuit takes the form~\eqref{circ:hardform}. The initial state $\ket{\phi_\text{init}}$ was chosen as the gauge invariant bare vacuum state with no particles on the sites and one unit of positive flux on the link.

The complete quantum circuit was run on the $ibmq\_vigo$ superconducting machine which is characterized in Fig.~\ref{fig:ibmq-vigo}. At each iteration step of the VQE algorithm, the parametrized  circuit was repeated 8000 times (8000 shots) in order to achieve a small sampling error for the energy and particle number expectation value. For the classical optimization routine, the \textit{SPSA} optimization algorithm (100 optimization steps, without calibration) was used which has been proposed for minimizing noisy energy functions~\cite{Kandala2017}. Furthermore, we applied a measurement-error mitigation scheme that is included in IBM's $\mathtt{Qiskit}$ software-development framework in order to reduce noise errors due to the measurement readout process~\cite{Qiskit}.

%\clearpage
\section{Gauge invariant variational form: Circuit example }\label{app:hardform-circuit}

Here, we display the explicit quantum circuit representing the variational form~\eqref{eq:unitary-varform-1} with 3 variational parameters $(\lambda_0,\, \lambda_1,\,\theta_0)$ that was used in the VQE simulations for the lattice QED model~\eqref{eq:1d-qed-hamiltonian} on a $d=1$ dimensional lattice with $M=2$ sites and with a spin value $S=1/2$ in the quantum link model truncation of the gauge electric fields.

The first two $X$ gates before the first barrier generate the gauge-invariant bare vacuum state with zero particles on the sites and positive flux on the link in between. The following variational form visibly consists of four Pauli strings $I_0X_1Y_2I_3Y_4\,,\ I_0X_1X_2I_3X_4\,,\ I_0Y_1Y_2I_3X_4\,,$ and $I_0Y_1X_2I_3Y_4$ which all commute with each other, thus implying a vanishing Trotter approximation error.
\begin{widetext}
\vspace*{0.25cm}
\begin{equation}\label{circ:hardform}
    \hspace*{2.3cm}
    \Qcircuit @C=1.0em @R=0.0em @!R {
	 	\lstick{ {\text{fermionic}}_{0} : \ket{0} } & \gate{X} \barrier[0em]{4} & \qw & \qw & \qw & \qw & \qw & \qw & \qw & \qw & \qw \\
	 	\lstick{ {\text{fermionic}}_{1} : \ket{0} } & \qw & \qw & \gate{H} & \ctrl{1} & \qw & \qw & \qw & \ctrl{1} & \qw & \qw \\
	 	\lstick{ {\text{fermionic}}_{2} : \ket{0} } & \gate{X} & \qw  & \gate{R_x(\frac{\pi}{2})} & \targ & \ctrl{2} & \qw & \ctrl{2} & \targ & \gate{R_x(-\frac{\pi}{2})} & \qw \\
	 	\lstick{ {\text{fermionic}}_{3} : \ket{0} } & \qw & \qw & \gate{R_z(\lambda_1)}  & \qw & \qw & \qw & \qw & \qw & \qw & \qw \\
	 	\lstick{ {\text{spin}}_{0} : \ket{0} } & \qw & \qw & \gate{R_x(\frac{\pi}{2})} & \qw & \targ & \gate{R_z\big(-\frac{\sqrt{3}}{3}\,\theta_0\big)} & \targ & \gate{R_x(-\frac{\pi}{2})} & \gate{H} & \qw \\
	 }
\end{equation}
\vspace*{0.2cm}
\begin{equation*}
    \Qcircuit @C=1.0em @R=0.0em @!R {
        \lstick{} & \qw & \qw & \qw & \qw & \qw & \qw & \qw & \qw & \qw & \qw & \qw   & \qw & \qw & \qw \\
        \lstick{} & \qw & \ctrl{1} & \qw & \qw & \qw & \ctrl{1} & \gate{H} & \gate{R_x(\frac{\pi}{2})} & \ctrl{1} & \qw & \qw  & \qw & \ctrl{1} & \qw \\
        \lstick{} & \gate{H} & \targ & \ctrl{2} & \qw & \ctrl{2} & \targ & \gate{H} & \gate{R_x(\frac{\pi}{2})} & \targ & \ctrl{2} & \qw & \ctrl{2} & \targ & \qw \\
        \lstick{} & \qw & \qw & \qw & \qw & \qw & \qw & \qw & \qw & \qw & \qw & \qw  & \qw & \qw & \qw \\
        \lstick{} & \qw & \qw & \targ & \gate{R_z\big(-\frac{\sqrt{3}}{3}\,\theta_0\big)} & \targ & \qw & \qw & \qw & \qw & \targ & \gate{R_z\big(-\frac{\sqrt{3}}{3}\,\theta_0\big)} & \targ & \gate{H} & \qw \\
    }
\end{equation*}
\vspace*{0.2cm}
\begin{equation*}
    \Qcircuit @C=1.0em @R=0.0em @!R {
        \lstick{} & \qw & \qw & \qw & \qw & \qw & \qw & \qw & \qw & \qw \barrier[0em]{4} & \qw  & \qw \\
        \lstick{} & \qw & \qw & \ctrl{1} & \qw & \qw & \qw & \ctrl{1} & \gate{R_x(-\frac{\pi}{2})} & \gate{R_z(\lambda_0)} & \qw & \qw \\
        \lstick{} & \gate{R_x(-\frac{\pi}{2})} & \gate{H} & \targ & \ctrl{2} & \qw & \ctrl{2} & \targ & \gate{H} & \qw & \qw & \qw \\
        \lstick{} & \qw & \qw & \qw & \qw & \qw & \qw & \qw & \qw & \qw & \qw & \qw \\
        \lstick{} & \gate{R_x(\frac{\pi}{2})} & \qw & \qw & \targ & \gate{R_z\big(\frac{\sqrt{3}}{3}\,\theta_0\big)} & \targ & \gate{R_x(-\frac{\pi}{2})} & \qw & \qw & \qw & \qw \\
    }
\end{equation*}
\end{widetext}


%\bibliographystyle{elsarticle-num}
%\clearpage

\end{document}